\documentclass[10pt,twocolumn,preprintnumbers,amsmath,amssymb,nofootinbib,superscriptaddress]{revtex4-1}

\usepackage{graphicx}
\usepackage{dcolumn}
\usepackage{amssymb,amsmath,bm}
\usepackage{color}
\usepackage[colorlinks,linkcolor=red,citecolor=blue,urlcolor=blue ]{hyperref}
\usepackage{multirow}
\usepackage{enumitem}
\usepackage{mathptmx} 
\usepackage[utf8]{inputenc}
\usepackage{tensor}
\usepackage{amssymb} 
\usepackage{lettrine}
\usepackage[normalem]{ulem}
\usepackage{empheq}
\usepackage{comment}

\def \be {\begin{equation}}
\def \ee {\end{equation}}

\newcommand{\beq}{\begin{equation}}
\newcommand{\eeq}{\end{equation}}
\newcommand{\bea}{\begin{eqnarray}}
\newcommand{\eea}{\end{eqnarray}}


\usepackage[stable]{footmisc}

\newcommand\ees{\end{eqnarray}}
\newcommand\bees{\begin{eqnarray}}

\begin{document}
  \title{Generation and propagation of nonlinear quasi-normal modes of a Schwarzschild black hole}

 \author{Macarena Lagos}
 \email{m.lagos@columbia.edu}
  \affiliation{Center for Theoretical Physics, Departments of Physics and Astronomy, Columbia University, New York, NY 10027, USA}
   \author{Lam Hui}
  \email{lh399@columbia.edu}
  \affiliation{Center for Theoretical Physics, Departments of Physics and Astronomy, Columbia University, New York, NY 10027, USA}

\begin{abstract}
In the analysis of a binary black hole coalescence, it is necessary to include gravitational self-interactions in order to describe the transition of the gravitational wave signal from the merger to the ringdown stage. In this paper we study the phenomenology of the generation and propagation of nonlinearities in the ringdown of a Schwarzschild black hole, using second-order perturbation theory. 
Following earlier work, we show that the Green's function and its causal structure determines how both first-order and second-order perturbations are generated, and hence highlight that both of these solutions share some physical properties.
In particular, we discuss the sense in which both linear and quadratic quasi-normal modes (QNMs) are generated in the vicinity of the peak of the gravitational potential barrier (loosely referred to as the light ring).
Among the second-order perturbations, there are solutions with linear QNM frequencies (whose amplitudes are thus renormalized from their linear values), as well as quadratic QNM frequencies with a distinct spectrum.  Moreover, we show using a WKB analysis that, in the eikonal limit, waves generated inside the light ring propagate towards the black hole horizon, and only waves generated
outside propagate towards an asymptotic observer. These results might be relevant for recent discussions on the validity of perturbation theory close to the merger. 
Finally, we argue that even if nonlinearities are small, quadratic QNMs may be detectable and would likely be useful for improving ringdown models of higher angular harmonics and  future tests of gravity.
\end{abstract}

  \maketitle

\section{Introduction}\label{sec:intro}
Coalescing black hole (BH) binaries emit gravitational waves (GWs) that allow us to probe gravity in the strong-field regime. These GWs are typically analyzed with different methods depending on the stage of the coalescence process. Initially, during the inspiral phase, when the black holes have small velocities compared to that of light, GWs can be studied analytically via the post-Newtonian formalism. Near the moment of the merger, GWs are sensitive to non-linear gravitational effects which are analyzed performing numerical relativity (NR) simulations. After the merger---in the ringdown phase---the coalescence process has culminated into a single perturbed black hole, whose GWs can be analyzed using black hole perturbation theory. 

In particular, during the ringdown, GWs are described by a linear superposition of quasi-normal modes (QNMs), which correspond to the resonant exponentially-decaying modes of the final black hole as it settles down to a stationary state. These modes have an infinite discrete spectrum of complex frequencies, $\omega=\omega_R+i\omega_I$, whose real part $\omega_R$ determines the oscillation timescale of the modes, whereas the imaginary part $\omega_I$ determines their exponential damping timescale (see e.g.\ \cite{Berti:2009kk} for a review on QNMs). 

In General Relativity (GR), the amplitude of each QNM depends on the initial conditions that led to the formation of the final black hole, but the QNM frequencies are universal since they are characterized solely by the mass, $M$, and angular momentum, $J$, of the final black hole. The QNM frequencies are labelled by three discrete numbers: the angular harmonic indices $(\ell,m)$ and the degree of the harmonic overtone number $n$. If there were additional fundamental fields present in the universe, they could affect the QNM spectrum of BHs and introduce new parameters determining the frequencies $\omega$. Therefore, the observation of QNM frequencies can be a powerful tool to test the properties of gravity (see e.g.\ \cite{Barausse:2008xv, Cardoso:2016ryw,Tattersall:2017erk, Berti:2018vdi,Franciolini:2018uyq,Hui:2021cpm}) and perform consistency tests of GR \cite{LIGOScientific:2020tif}. 

As previously mentioned, the merger process is believed to be highly non-linear. However, since the QNMs decay exponentially fast in time, at some time $t_\text{ref}$ after the merger, nonlinearities are expected to become irrelevant and the QNMs can be analyzed using \emph{linear} perturbation theory. Nevertheless, there has been some debate concerning the optimal choice of $ t_\text{ref}$
(see related discussions in e.g.\ \cite{Berti:2007fi, Bhagwat:2017tkm}), given the fact that if chosen too late then there will not be enough ringdown signal left in the available data due to its fast decay, and if chosen too early then contamination from nonlinearities may bias the linear analysis. This issue raises the crucial questions of how close to the merger linear theory can describe well the GW signal, and what the relevance of nonlinearities is. In this paper, we make some preliminary steps
in this direction by understanding the phenomenological properties of the generation and propagation of second-order BH perturbations.
The hope is that this will help improve ringdown models, and enable
the optimal analysis of high quality GW data expected in the future. In particular, the inclusion of nonlinearities in ringdown models will potentially allow for unbiased constraints of quasi-normal modes, and thus more confident tests of gravity. In addition, the detection of nonlinearities would allow to test the nonlinear dynamical predictions of GR.

So far, numerical studies have obtained varied conclusions on the relevance of nonlinearities. While it has been known for some time that the inclusion of linear overtones in ringdown models improve the fits to GW waveforms (see e.g.\ \cite{Baibhav:2017jhs}), \cite{Buonanno:2006ui, Giesler:2019uxc} confirmed that linear QNMs with up to 7 overtones fit well NR simulations of the $(\ell=2, |m|=2)$ GW signal from non-precessing nearly-equal mass binary black hole (BBH) mergers, {\it all the way back to the moment of the merger}, or even slightly before. These analyses assumed that the QNM frequencies were given by the predictions from linear BH perturbation theory in GR, and fit for their amplitudes since these cannot be easily predicted due to their dependence on pre-merger history.
Subsequent numerical analyses have included higher harmonics, and confirmed that a similar linear ringdown analysis can indeed fit well waveforms of various binary BH systems \cite{Bhagwat:2019dtm, Cook:2020otn, JimenezForteza:2020cve, Dhani:2020nik, MaganaZertuche:2021syq}. These results are somewhat surprising since the physics of the merger is expected to be highly non-linear. For instance, \cite{Finch:2021iip} concludes that for precessing binary systems, linear QNMs do not always fit well GW signals from NR simulations starting from the merger time. Nevertheless, these results have motivated the use of the entire post-merger signal of current GW events, such as GW150914 \cite{LIGOScientific:2016vbw}, to detect the fundamental QNM ($n=0$) as well as the first overtone ($n=1$), and to perform tests of gravity \cite{LIGOScientific:2020tif, Isi:2019aib, Isi:2020tac, Capano:2021etf}, although different conclusions have been obtained \cite{Bustillo:2020buq, Cotesta:2022pci, Isi:2022mhy, Finch:2022ynt}.

In this paper, we adopt an analytic approach to nonlinearities, making use of
black hole perturbation theory to second order. A particular focus, though not
an exclusive one, will be on the quadratic QNMs (here dubbed QQNMs).
There have been a number of investigations on this topic, starting with analyses on Schwarzschild black holes \cite{Gleiser:1995gx, Gleiser:1998rw, Garat:1999vr, Brizuela:2006ne, Brizuela:2007zza, Ioka:2007ak, Nakano:2007cj, Okuzumi:2008ej, Brizuela:2009qd}, which characterized the QQNM frequency spectrum and the sources that drive these quadratic modes, followed by generalizations to Kerr black holes \cite{Campanelli:1998jv, Green:2019nam, Loutrel:2020wbw, Ripley:2020xby}. Our goal in this paper is
to understand better how, when and where the second-order perturbations, in particular the QQNMs, are generated, and how they propagate locally. For simplicity, our investigation is confined to perturbations around a Schwarzschild black hole, though some of the conclusions are expected to translate straightforwardly to a Kerr black hole.  Black hole perturbation theory up to second order
has the following schematic form:
\begin{eqnarray}
\label{schematic}
{\cal D} h^{(1)} \sim 0 \quad , \quad {\cal D} h^{(2)} \sim h^{(1)} {}^2 \, .
\end{eqnarray}
The first equation is linear perturbation theory: $h^{(1)}$ is the first-order
metric perturbation (indices suppressed) around the black hole, and ${\cal D}$ is a linear differential operator which
contains up to two derivatives, and has a non-trivial effective gravitational potential. 
The second equation shows how the second-order
perturbation $h^{(2)}$ is sourced by quadratic combinations of $h^{(1)}$ 
(with derivatives acting on $h^{(1)}$ kept implicit). Importantly, the {\it same} operator ${\cal D}$ 
appears in both equations. Our focus in this paper is not on the detailed form
of the $h^{(1)} {}^2$ terms on the right hand side; they have been worked out
in pioneering papers by \cite{Gleiser:1995gx, Gleiser:1998rw}, and we will
make use of certain general features of their results.
Rather, our goal is to study the implications of the operator ${\cal D}$
for the generation and propagation of the second-order perturbations.

Among our findings, let us highlight several key points, some of which are known from  earlier analyses.

1. Given a pair of modes from the linear QNM frequency spectrum 
$\omega^{(1)}=\omega^{(1)}_R+i\omega^{(1)}_I$ and
$\omega^{(1)'}=\omega^{(1)'}_R+i\omega^{(1)'}_I$, one can see from Eq.\ (\ref{schematic}) that they will generate a quadratic QNM frequency
$\omega^{(2)}=\omega^{(2)}_R+i\omega^{(2)}_I$ given by
$\omega^{(2)}_{R}=\omega^{(1)}_{R} \pm \omega^{(1)'}_{R}$ and $\omega^{(2)}_I=\omega^{(1)}_I+\omega^{(1)'}_I$  \cite{Gleiser:1995gx,Ioka:2007ak,Nakano:2007cj}. This means that there is a new distinct quadratic frequency spectrum of QNMs, which is fixed and constructed from linear QNM frequencies.

2. We formalize the above intuition using the Green's function approach, which
provides further insights. We find that the second-order solution is in general
a superposition of modes with the quadratic QNM spectrum $\omega^{(2)}$ (as shown in \cite{Okuzumi:2008ej}), 
{\it and} modes with the linear QNM spectrum $\omega^{(1)}$
\footnote{The second-order solution also 
has parts that are unrelated to QNMs or QQNMs, such as polynomial tails \cite{Okuzumi:2008ej}. See further discussion below.}.
This is in agreement with previous numerical results \cite{ Pazos:2010xf, Ripley:2020xby}.
Importantly, this result means that the net amplitude of modes with linear frequencies $\omega^{(1)}$ receive a nonlinear renormalization.

3. The Green's function's causal structure sheds light
on the times and locations of linear and quadratic QNM generation.
The amplitudes of the QQNMs depend on signals that have enough
time to reach the light ring
\footnote{We use the term light ring loosely to refer to the location
of the top of the potential in the operator ${\cal D}$ in Eq.\ (\ref{schematic}).}
and then the observer (analogous to previous results for linear QNMs \cite{Andersson:1996cm, Szpak:2004sf}). This supports the build-up picture in which the QNM amplitudes may accumulate over time as more
of the initial perturbations become causally connected to the
observer and the light ring; the amplitudes of QNMs are
in general not constant at all times {\it even within linear theory}.

4. To gain a better understanding of how the different parts of the Green's function
dictate both the linear evolution and the generation of second-order perturbations,
we work out a simple toy problem: that of a delta function
potential. The solution can be written down in closed form,
and illustrates explicitly the key results outlined above.

5. We use the Wentzel-Kramers-Brillouin (WKB) approach to study the QNM local propagation in the high-frequency limit. We show that both linear and quadratic QNMs generated near the horizon propagate towards the black hole, whereas only those generated outside the light ring of the black hole will propagate to the observer. This result analytically confirms that not all of the GWs escape to infinity, as part of them are swallowed by the black hole. A related result was found recently in toy simulations in \cite{Sberna:2021eui}, where absorption of the initial QNM signal led to an increase of the black hole horizon. This result is important to take into account, given that previous NR simulations find large perturbations right after merger to be generally confined to regions
very close to the black hole horizon \cite{Bhagwat:2017tkm, Okounkova:2020vwu}, which lends some credence to the notion that while large perturbations
exist very close to the horizon right after merger, the observable QNMs asymptotically far are not necessarily sensitive to them.
This idea has been conjectured by some authors \cite{Gleiser:1995gx,Okounkova:2020vwu} in the past. 

6. At a practical level, including QQNMs in ringdown
waveform analyses of simulations and data should prove
beneficial. Previous analyses of head-on black hole collisions have shown model
improvement when including second-order perturbations 
\cite{Gleiser_1996, Nicasio_1999}. 
In this paper, we discuss when the amplitude of nonlinearities is large enough to be relevant in ringdown models. We show that the answer depends strongly on the angular harmonic structure of the signal.
Take for example a nearly equal-mass binary merger. At the linear level, the amplitude is dominated by the $(\ell=2, |m|=2)$ angular mode, with subdominant higher harmonics
(see e.g.\ \cite{Kamaretsos:2011um,Ota:2019bzl,JimenezForteza:2020cve}). 
At second order, one then expects the largest quadratic QNM mode to have $(\ell=4,|m|=4)$, originating from
the product of two linear $(\ell=2, |m|=2)$ modes. We make a simple dimensional analysis to conclude that its amplitude can be comparable to or larger
than that of the linear QNM $(\ell=4,|m|=4)$.

From these results, we conclude that nonlinear QNMs are expected to always be generated after the merger. Nonetheless, analytical models that only assume the presence of linear QNMs frequencies may work better than expected because: (i) nonlinear effects are partially included in those models through their renormalized amplitudes, and (ii) the signal generated close to the horizon, which is expected to contain the most amount of nonlinearities, will not propagate to asymptotic observers. 

In addition, the amplitude of nonlinearities highly depend on the angular harmonic structure of the signal. Previous works using linear QNMs to model the merger \cite{Giesler:2019uxc, Bhagwat:2019dtm} focused on $(\ell=2, |m|=2)$ harmonics which, based on dimensional estimations, are expected to have sub-percent level corrections from nonlinearities for a nearly equal-mass quasi-circular binary black hole coalescence (see Appendix \ref{app:CB}). Instead, as previously mentioned, $(\ell=4, |m|=4)$ harmonics could have large contributions from nonlinearities. This appears to be the case in the numerical analysis of \cite{London:2014cma}, and has been confirmed as well in the recent works developed in parallel to this paper \cite{Mitman:2022qdl,Cheung:2022rbm}.
Therefore, future analyses must be careful when using linear QNMs frequencies to describe higher harmonics. Indeed, the recent study in \cite{Ma:2022wpv} has also shown evidence of quadratic QNMs in $(\ell=5, |m|=4)$ and $(\ell=5, |m|=5)$ harmonics in at least one specific binary merger simulation.
Furthermore, higher harmonics are expected to be important in future GW data. Already a recent analysis of the event GW190521 has claimed evidence for a sub-dominant higher harmonic $(\ell=3,|m|=3)$ \cite{Capano:2021etf}, and third-generation GW detectors could observe between $10^2-10^4$ events with detectable higher harmonics in the ringdown \cite{Berti:2016lat, Ota:2019bzl, Bhagwat:2021kwv}. 
In addition, LISA will observe supermassive black holes binaries with mass $M>10^6 M_\odot$, where most of the signal will come from the ringdown since they will have no (or little) detectable inspiral signal due to its low frequency. In these cases, the analysis of higher harmonics will be crucial for extracting information about the progenitor's masses \cite{Kamaretsos:2011um} as well as the inclination, luminosity distance and localization of the source \cite{Baibhav:2020tma}. 

This paper is organized as follows. In Section \ref{QNM_freq} we review the general setup for
second-order perturbations around a Schwarzschild black hole, discussing their angular, radial and temporal structures using separation of variables.
In Section \ref{sec:greens} we use the Green's function approach to  confirm and generalize previous findings on the temporal and angular profiles of second-order perturbations, and we work through a toy model to illustrate important features about the linear and quadratic QNMs, as well as the role of causality. 
In Section \ref{QNM_space} we analyze the radial profile of the QQNMs in the eikonal limit, which determines the propagation direction of GWs. We consider both near  horizon and spatial infinity regimes using the WKB formalism. In Section \ref{energy} we discuss the relevance of QQNMs with a simple dimensional analysis, and conclude in  Section \ref{conclusions} with a  summary and discussion of our findings. 

We set the speed of light to unity in this paper. Since we make use of a number of analytical techniques to analyze the behavior of quadratic QNMs, to ease readability, we compile common symbols used throughout this paper in Table \ref{table:symbols}, indicating the location where they were defined for the first time, and their meaning. 
\begin{widetext}

\begin{table}[h!]
\centering
\begin{tabular}{ |c|c|c| } 
 \hline
 Notation & Equation & Meaning \\ [0.5ex]  
 \hline\hline
 $\epsilon$ & Eq.\ (\ref{h_expansion}) & Expansion parameter in the metric amplitude  \\
\hline
 $\xi=1/\ell$ & Above Eq.\ (\ref{Linear_ring_away}) & Expansion parameter in the angular harmonic number $\ell$ \\
 \hline
 $\delta=GM/r$ & Above Eq.\ (\ref{delta_exp}) & Expansion parameter in the radial distance from the source\\
 \hline
  $\Delta r = (r_*-\hat{r}_*)/(MG)$ & Below Eq.\ (\ref{W_Taylor}) & Expansion parameter in the radial distance from light ring location $\hat{r}_*$\\
  \hline
  $z\sim \Delta r/\sqrt{\xi}$ &  Eq.\ (\ref{Linear_ring_sol}) & Suitable radial variable such that $z\rightarrow \infty$ describes eikonal limit \\
 \hline
 $X^{(n)}$ & Eq.\ (\ref{h_expansion}) & $\epsilon^n$ order contribution to a variable $X$\\
  \hline
   $X_{\xi n}$ & Eq.\ (\ref{U_expanded}) & $\xi^n$ order contribution to a variable $X$\\
   \hline
    $X_{ij}$ & Eq.\   (\ref{u_radial_exp}) & $\xi^i \Delta r^j$ order contribution to a variable $X$ \\
  \hline
   ${}_{s}Y_{\ell m}(\theta,\phi)$ & Eq.\ (\ref{harmonic_product}) & spin $s$-weighted $(\ell, m)$ spherical harmonic \\
   \hline
   $\omega_R$, $\omega_I$ & Above Eq.\ (\ref{QQNM_frequency}) & Real and imaginary parts of any QNM frequency\\
\hline
$\omega_{\pm}$ & Eq.\ (\ref{S2_gral}) & Quadratic QNM frequencies constructed from the sum or  (conjugated) difference of linear QNMs\\
\hline
 ${}^{e,o}\Psi$ & Eqs.\ (\ref{Psi_Eqlinear})-(\ref{Q_Eqlinear}) & even (Zerilli) and odd (Regge-Wheeler) radial variables\\
  \hline
   $\Psi_F$, $\Psi_Q$, $\Psi_B$ & Eqs.\ (\ref{Psi1_Green_split}) & even/odd variables from the green's function pieces $G_F$, $G_Q$ and $G_B$ described in Subsec.\ \ref{setup} \\
  \hline
  $V_Z$, $V_{RW}$ & Eqs.\ (\ref{VZ})-(\ref{VRW}) &  Zerilli and Regge-Wheeler radial potentials\\
 \hline
  $U=\omega^2-V$ & Eq.\ (\ref{Linear_WKB}) &  Effective potential $U$ for  Regge-Wheeler ($V=V_{RW}$)  and Zerilli ($V=V_Z$) variables\\
 \hline
\end{tabular}
\caption{Summary of notation used throughout this paper, location where it was introduced, and  associated meaning.}\label{table:symbols}
\end{table}
\end{widetext}

\section{Second-order Perturbations and Quadratic QNMs --- General Setup}\label{QNM_freq}
Let us start by considering perturbations of the spacetime metric $g_{\mu\nu}$ as:
\begin{align}
    g_{\mu\nu}=\bar{g}_{\mu\nu}+ h_{\mu\nu}; \quad h_{\mu\nu} \equiv \epsilon h^{(1)}_{\mu\nu} + \epsilon^2 h^{(2)}_{\mu\nu}+\mathcal{O}(\epsilon^3), \label{h_expansion}
\end{align}
where $\epsilon \ll 1$ is the perturbation theory parameter and $h^{(j)}_{\mu\nu}$ is the $j$-th order perturbation around the background $\bar{g}_{\mu\nu}$. For simplicity, in this paper we assume the background to be given by an isolated Schwarzschild black hole:
\begin{equation}
    d\bar{s}^2= -f(r) dt^2 + f(r)^{-1}dr^2 + r^2(d\theta^2+ \sin(\theta)^2d\phi^2),\label{Sch_bkgd}
\end{equation}
where $f(r)=1-r_s/r$ and $r_s=2GM$ is the Schwarzschild radius, with
$M$ the mass of the black hole and $G$ the gravitational constant. 
The Einstein equations in vacuum can be Taylor expanded in the parameter $\epsilon$ and be schematically expressed as:
\begin{align}
    G_{\mu\nu}(g)=& G_{\mu\nu}^{(0)}(\bar{g})+ \epsilon G_{\mu\nu}^{(1)}(h^{(1)})\nonumber\\
    &+ \epsilon^2 \left[G_{\mu\nu}^{(1)}(h^{(2)})+ G^{(2)}_{\mu\nu}(h^{(1)},h^{(1)})\right]+ \mathcal{O}(\epsilon^3)=0,
\end{align}
where $G_{\mu\nu}$ is the Einstein tensor, and $G_{\mu\nu}^{(j)}$ indicates its $j$-th order Taylor expansion in the perturbation $h_{\mu\nu}$. This equation is satisfied when each $\epsilon^n$ contribution vanishes separately. At leading order, we have $G_{\mu\nu}^{(0)}(\bar{g})=0$ which is the background equation of motion, a solution of which is Eq.\ (\ref{Sch_bkgd}). 
At first and second order in $\epsilon$, we have:
\begin{align}
    & G_{\mu\nu}^{(1)}\left(h^{(1)}\right)=0,\label{lin_EoM}\\
    & G_{\mu\nu}^{(1)}\left(h^{(2)}\right)= - G^{(2)}_{\mu\nu}\left(h^{(1)},h^{(1)}\right) \equiv S_{\mu\nu}^{(2)}.\label{Quad_EoM}
\end{align}
From these results it is clear that the second-order equation of motion (\ref{Quad_EoM}) has the same left-hand side (LHS) structure as the first-order one, but it has an effective source term $S_{\mu\nu}^{(2)}$ determined by the quadratic product of the first-order metric perturbations $h^{(1)}$. This source will induce non-trivial particular solutions to Eq.\ (\ref{Quad_EoM}), which will determine the spectrum of the QQNMs\footnote{Note that the homogeneous solution to Eq.\ (\ref{Quad_EoM}) will not be considered part of the QQNMs spectrum here since it will instead have the same linear QNMs frequencies as the first-order perturbations.}.

Before we proceed further, let's clarify perhaps a pedantic point.
The definition of $\epsilon$, the perturbation expansion
parameter, is location dependent. For instance, at the location of a far
away observer, the expected metric perturbations are extremely small (for
instance, typical GW strain is at the $10^{-22}$ level) and
thus linear perturbation theory, essentially around Minkowski space,
is highly accurate at the observer. On the other hand, the metric
perturbations close to the black hole are considerably larger, and
the expansion parameter $\epsilon$ should be understood to be defined
in that neighborhood. As far as the asymptotic observer is concerned,
the detailed dynamics close to the black hole generates
$h_{\mu\nu}^{(1)}$ and $h_{\mu\nu}^{(2)}$, and both fall off
inversely proportional to distance, far enough away from the black
hole (see further discussion in Sec. \ref{energy}).
Previous authors have estimated that the QQNMs could give a correction
of about $10\%$ to the linear QNMs at the detector \cite{Ioka:2007ak,Nakano:2007cj}.

The (real) metric perturbation at each order can be written as
the real portion of its complex counterpart:
\begin{equation}
h^{(j)}_{\mu\nu} = {\,\rm Re\,} \left( h^{c(j)}_{\mu\nu}\right) \, .
\end{equation}
As such, Eqs.\ (\ref{lin_EoM}) and (\ref{Quad_EoM}) can be recast
as (Appendix \ref{app:complex}):
\begin{align}
    & G_{\mu\nu}^{(1)}\left(h^{c(1)}\right)=0,\label{linEqn_complex}\\
    & G_{\mu\nu}^{(1)}\left(h^{c(2)}\right)= - G^{(2)}_{\mu\nu}\left(\frac{1}{2}(h^{c(1)}+h^{c(1)*}),h^{c(1)}\right)
      \, .\label{QEqn_complex}
\end{align}
Performing a separation of variables, we can write
\begin{align}\label{Perts_Ansatz}
    h^{c(j)}_{\mu\nu} = \int {d\omega \over 2\pi} \sum_{a=1}^{10} \sum_{\ell, m}
  H^{a(j)}_{\ell m \omega}(r)\; e^{-i\omega t}\; T_{\ell
  m;\mu\nu}^a(\theta,\phi) \, ,
\end{align}
where $H^{a(j)}_{\ell m \omega}$ is the radial function of the $j$-th
order metric perturbation for each tensor spherical harmonic
$T^a_{\ell m; \mu\nu}$ (labeled by $a$ from 1 to 10 accounting for the
10 different metric components)
\cite{PhysRev.108.1063, PhysRevD.2.2141}.

In the rest of this section, we highlight several broad features of
Eqs.\ (\ref{linEqn_complex}) and (\ref{QEqn_complex})
that are relevant for our goal
of understanding the generation and propagation of nonlinearities.
The discussion will be schematic, since the details are not important
for our purpose. The reader is referred to 
\cite{Brizuela:2006ne,Brizuela:2009qd} for further discussions.

\subsection{Angular structure}\label{sec:angle_time}

Imagine plugging Eq.\ (\ref{Perts_Ansatz}) into
Eq.\ (\ref{QEqn_complex}). We see that a product
of angular harmonics on the right gives rise to
a sum of angular harmonics on the left.
Specifically, in a Schwarzschild background,
the angular tensors $T^{a}_{\ell m}$ are constructed from spherical
harmonics $Y_{\ell m}(\theta,\phi)$ and their derivatives as in
\cite{PhysRevD.2.2141} or, equivalently, from spin-weighted spherical
harmonics ${}_{s}Y_{\ell m}(\theta,\phi)$ \cite{1967JMP.....8.2155G}
(which are defined when $|s|\leq \ell$ and $|m|\leq \ell$). 
The product of two spin-weighted spherical harmonics can be
re-expressed as a linear superposition of spin-weighted spherical
harmonics---this is why we use the same 
angular decomposition in Eq.\ (\ref{Perts_Ansatz}) for linear and
second (and higher) order perturbations. 
In other words, we use the following property of spin-weighted
spherical harmonics, which form a complete and orthonormal set \cite{1967JMP.....8.2155G}
\begin{align}\label{harmonic_product}
&\sum_{\ell_2 m_2} \frac{k(\ell)k(\ell')}{k(\ell_2)} 
C(\ell,m,\ell',m';\ell_2, m_2)C(\ell,s,\ell',s';\ell_2,
  s_2) \times \nonumber \\
&\quad \quad \;{}_{s_2}Y_{\ell_2 m_2}(\theta,\phi) 
= {}_{s}Y_{\ell m}(\theta,\phi){}_{s'}Y_{\ell' m'}(\theta,\phi),
\end{align}
where $k(\ell)=\sqrt{2\ell+1}/\sqrt{4\pi}$, and $C$'s are the
Clebsch-Gordan coefficients that are non-vanishing only if $s_2=s+s'$,
$m_2=m+m'$ and $|\ell-\ell'|\leq \ell_2\leq |\ell+\ell'|$.
This expression helps determine the angular structure of second-order perturbations in terms of that of the first-order perturbations.
Note that because of the relationship $|\ell-\ell'|\leq
\ell_2\leq|\ell+\ell'|$, the second-order perturbations will generally
have non-vanishing propagating modes with $\ell_2<2$, contrary to the
linear propagating modes, which must have $\ell , \ell' \ge
2$. However, in the large radius $r$ limit, only the spin $s=-2$
spherical harmonics are relevant (due to the peeling theorem
\cite{Sachs:1961zz,Sachs:1962wk, Newman:1961qr}) and thus modes with
$\ell_2=0,1$ are not observationally relevant.

In addition, note that a given spherical harmonic of the
second-order perturbations can be sourced by various multiplications
of the linear ones. For instance, a second-order $(\ell_2 = 4, m_2 = 4)$ can be sourced by
the linear $(\ell = 2, m = 2)\times (\ell' = 2, m' = 2)$, $(\ell = 3, m=2)\times(\ell'=2,m'=2)$, and so on.
In particular, for QNMs with their distinctive frequencies,
this means there are many quadratic QNM frequencies (an infinite
number in fact) associated with a given spherical harmonic,
similar to the way there are many overtones for linear QNMs of a given harmonic.

Furthermore, since the background is invariant under parity, 
it is useful to split the angular tensors and radial functions into parity even and parity odd parts, following Regge-Wheeler \cite{PhysRev.108.1063}.
The parity even modes transform as $(-1)^\ell$ while the parity odd
modes transform as $(-1)^{\ell+1}$. At the level of linear theory, the
two set of modes do not mix. At second order, it is still true the
second-order even modes and the second-order odd modes do not mix.
However, the second-order 
even modes can be generated from a number of sources: 
linear even $\times$ linear even, linear odd $\times$ linear odd,
{\it and} linear odd $\times$ linear even. (Likewise for the second-order odd modes.) There is a simple rule
governing the first and second-order perturbations in harmonic space
\cite{Brizuela:2006ne, Brizuela:2007zza, Brizuela:2009qd}:
\begin{equation}\label{Parity_Condition}
(-1)^{\ell_2}\sigma_2=(-1)^{\ell}(-1)^{\ell'}\sigma\sigma' \, ,
\end{equation}
where $\sigma$ and $\sigma'$ ($= \pm 1$) are the parity of the two
linear modes, and $\sigma_2$ is the parity of the second-order one.

\subsection{Radial structure}
\label{sec:QNM_Eqn}

Of the 10 metric components, there are 2 propagating
degrees of freedom. Regge and Wheeler \cite{PhysRev.108.1063}
and Zerilli \cite{PhysRevD.2.2141} showed how to isolate these
2 degrees of freedom in linear perturbation theory and obtain
equations of the form:
\begin{align}
    &
      \partial_{r_*}^2 \, {}^{e}\Psi^{(1)}(r_*)+\left(\omega^{2}-V_Z(r)\right){}^{e}\Psi^{(1)}(r_*)= 0, \label{Psi_Eqlinear}\\
    &
      \partial_{r_*}^2 \, {}^{o}\Psi^{(1)}(r_*)+\left(\omega^{2}-V_{RW}(r)\right){}^{o}\Psi^{(1)}(r_*)=0,
\label{Q_Eqlinear}
\end{align}
where ${}^{e}\Psi^{(1)}$ and ${}^{o}\Psi^{(1)}$ represent the 
Zerrilli (even) and Regge-Wheeler (odd) variables
(each formed from judicious combinations
of $H^{a(1)}_{\ell m \omega}$ defined in Eq.\ (\ref{Perts_Ansatz})). Here, 
$\partial_{r_*}$ denotes derivative with respect to the tortoise
coordinate: $r_* \equiv r + {\,\rm ln\,} (r/r_s - 1)$. 
These equations are written in frequency-angular-harmonic-space,
i.e. we are focusing on a mode with given $\omega, \ell, m$
(but suppressing the $\ell, m$ labels).
Keep in mind the most general solution involves a superposition
of the form (\ref{Perts_Ansatz}). 

It was further shown by \cite{Gleiser:1995gx, Gleiser:1998rw,
  Nakano:2007cj, Brizuela:2009qd}
that a second-order version of the Zerilli and Regge-Wheeler
variables can be defined, which obey:
\begin{align}
    & \partial_{r_*} \, {}^{e}\Psi^{(2)}(r_*)+\left(\omega^{2}-V_Z(r)\right){}^{e}\Psi^{(2)}(r_*)={}^{e}S^{(2)}(r_*), \label{Psi_Eq}\\
    &  \partial_{r_*} \,
      {}^{o}\Psi^{(2)}(r_*)+\left(\omega^{2}-V_{RW}(r)\right){}^{o}\Psi^{(2)}(r_*)={}^{o}S^{(2)}(r_*),\label{Q_Eq}
\end{align}
where ${}^{e}S^{(2)}$ and ${}^{o}S^{(2)}$ represent the sources for the second-order even and odd perturbations, respectively.
Each source consists of products of two first-order metric 
perturbations
and their derivatives, which can be reconstructed from 
${}^{o,e}\Psi^{(1)}$ \cite{Brizuela:2009qd}.
The reconstruction means the sources can be fully expressed in terms
of products of ${}^{o,e}\Psi^{(1)}$.
Some examples of quadratic sources in the Regge-Wheeler gauge can be
found in \cite{Gleiser:1998rw, Nakano:2007cj}, and a gauge-invariant
approach was studied in \cite{Brizuela:2009qd}
\footnote{We do not dwell on gauge issues here, since they have
been thoroughly discussed in \cite{Gleiser:1998rw, Brizuela:2007zza}. In broad stroke, they can be understood as
follows. At the linear level, we have schematically that $\tilde h^{(1)} \sim h^{(1)} + \xi^{(1)}$,
where $\xi^{(1)}$ represents a first-order coordinate transformation and
its derivatives (indices are suppressed; $\tilde h^{(1)}$ is the metric
perturbation in the new coordinates, while $h^{(1)}$ is the metric
perturbation in the old ones). Gauge fixing typically corresponds to
choosing $\xi^{(1)}$ such that certain components of $\tilde h^{(1)}$
vanish. The remaining non-vanishing components then represent the
desired physical degrees of freedom and auxiliary fields. Alternatively, one can use the gauge
choice to express $\xi^{(1)}$ in terms of $h^{(1)}$, and substitute
that into expressions for the non-vanishing components of $\tilde
h^{(1)}$, which can then be re-interpreted as gauge-invariant
combinations of components of $h^{(1)}$ (see
\cite{Franciolini:2018uyq} Appendix G for concrete examples).
At second order, we expect $\tilde h^{(2)} \sim h^{(2)} + \xi^{(2)} + 
\xi^{(1)2} + h^{(1)} \xi^{(1)}$ (where we have suppressed
derivatives and indices). The procedure for linear theory translates
straightforwardly to second order: gauge fixing means choosing
the appropriate coordinate transformation at second order $\xi^{(2)}$;
gauge-invariant combinations can be found in a similar way.
}.
Eqs.\ (\ref{Psi_Eq}) and (\ref{Q_Eq}) can be
generalized to higher orders \cite{Brizuela:2007zza}.

\begin{figure}[h!]
\centering
\includegraphics[width = 0.40\textwidth]{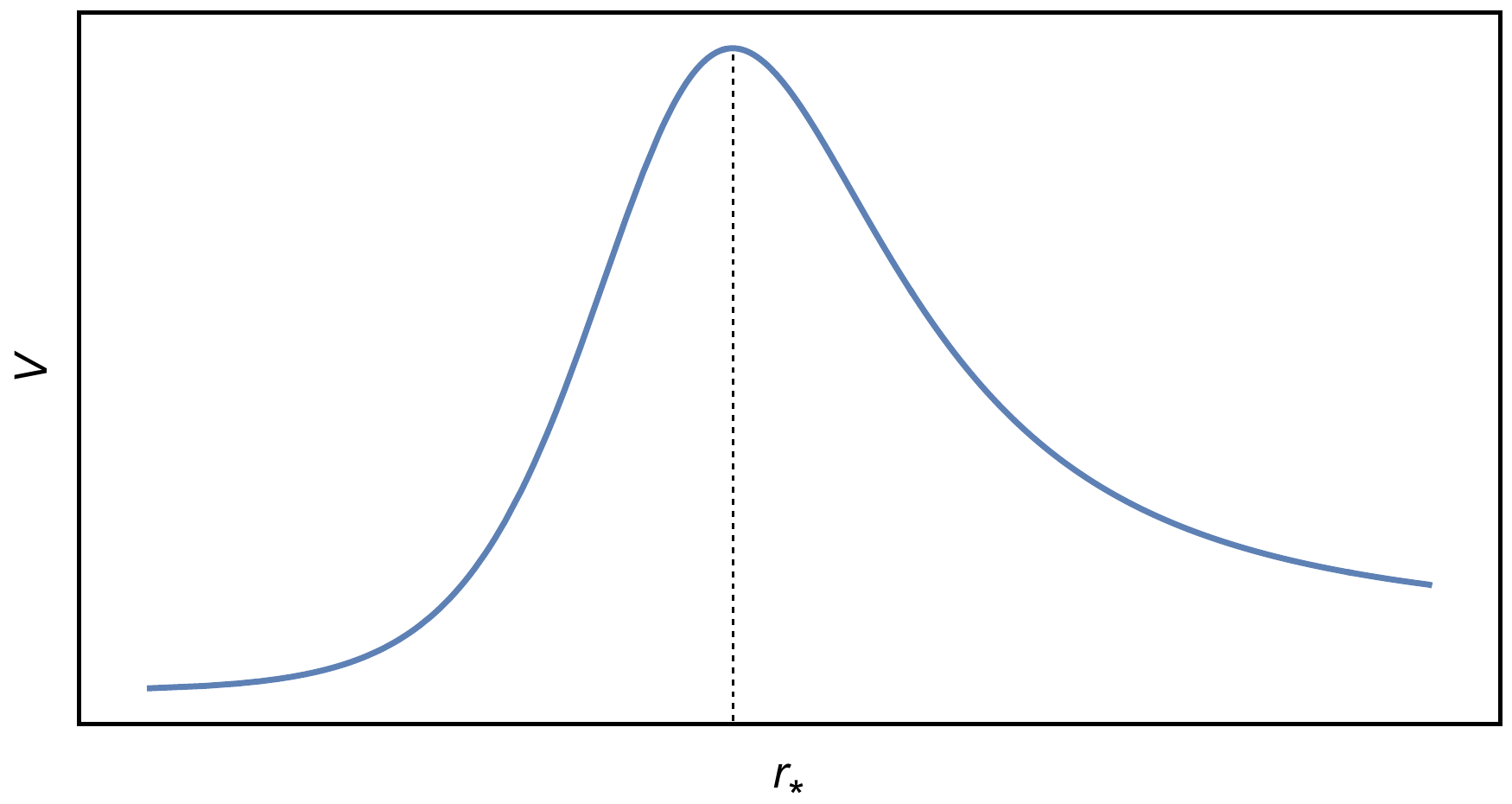}
 \caption{A schematic sketch of Regge-Wheeler/Zerilli potential as a
   function of the tortoise coordinate $r_*$. The horizon is at
   $r_*\rightarrow -\infty$ and spatial infinity at $r_*\rightarrow
   +\infty$, and the potential approaches zero in both limits. The
   potential has a maximum at a particular radius $r_{*}=\hat{r}_*$
   indicated by the vertical dashed line.}
 \label{potential_shape}
\end{figure}

The same potentials $V_Z$ and $V_{RW}$ show up
in both the first and second-order radial equations.
They are given by:
\begin{align}
    &V_Z(r) =2f(r)\frac{L^2r^2[(L+1)r+3GM]+9G^2M^2(Lr+GM)}{r^3(Lr+3GM)^2}, \label{VZ}\\
    & V_{RW}(r) = f(r)\left(\frac{\ell(\ell+1)}{r^2}-\frac{6GM}{r^3}\right),\label{VRW}
\end{align}
where $2L\equiv (\ell+2)(\ell-1)$. 
The Zerrilli and Regge-Wheeler potentials ($V_Z$ and $V_{RW}$) 
have the general radial shape shown in Fig.\ \ref{potential_shape}.
The potentials approach a constant (zero) near the horizon
($r_*\rightarrow -\infty$) and at spatial infinity ($r_*\rightarrow
+\infty$), and they reach a maximum at some special value $\hat{r}_*$,
which is $\ell$-dependent but approaches the light ring
$\hat{r}_*\rightarrow 3GM$ as $\ell\rightarrow \infty$. 
Throughout this paper, we use the term light ring to loosely refer to
the top of the potential, for any $\ell$.
Note that this general shape applies even for perturbations around a
Kerr black hole, if suitable variables are chosen, and the potential
will have $\omega$ and $m$ dependence \cite{1977RSPSA.352..381D, PhysRevD.41.374}.

It is worth stressing that there are many possible choices for
the second-order Regge-Wheeler/Zerilli variables.
One could redefine $\Psi^{(2)}$ (both even and odd) by 
adding extra terms that depend quadratically on the linear
perturbations---the resulting variables would still satisfy 
Eqs.\ (\ref{Psi_Eq})-(\ref{Q_Eq}) but with correspondingly different source terms. 
Following \cite{Gleiser:1995gx}, it is useful to take advantage of
this freedom, to modify the source terms so they have the desired
fall-off at large distances and close to the horizon, namely:
\begin{align}
    & {}^{e,o}S^{(2)}\sim {}^{e,o}\Psi^{(1)2} r^{-2} \quad \text{for}\quad r\rightarrow \infty,\label{Source_Infinity}\\
     & {}^{e,o}S^{(2)}\sim {}^{e,o}\Psi^{(1)2}(r-2GM) \quad \text{for}\quad r\rightarrow 2GM.\label{Source_Horizon}
\end{align}
Since the linear QNM solutions ${}^{e,o}\Psi^{(1)}$ behave as $\exp\{-i \omega(t\pm r_*)\}$ with constant amplitude in
the $r_* \rightarrow \pm \infty$ limit, the source terms
chosen thus have an analogous 
scaling with radius as the potentials in
Eqs.\ (\ref{VZ})-(\ref{VRW})\footnote{Keeping $t\pm r_*$ fixed.}. As we will see in Sec.\
\ref{QNM_space}, if the sources had a slower scaling with radius than
the above (at the
horizon or infinity), then the solutions for the quadratic QNMs
${}^{o,e}\Psi^{(2)}$ would have a divergent power-law
scaling. On the other hand, if the sources decayed faster than this, the
quadratic QNMs would have an asymptotically vanishing scaling at
the horizon and infinity. Ultimately, the physics is independent of the
choice of the second-order variables, but the choice of 
(\ref{Source_Infinity})-(\ref{Source_Horizon}) helps give the linear
and quadratic QNMs the same asymptotic behavior and a direct relation to physical quantities such as energy radiated.

\subsection{Temporal structure: QNMs}

The (linear) Regge-Wheeler (\ref{Q_Eqlinear})
and Zerrilli (\ref{Psi_Eqlinear}) equations are typically solved with the boundary
conditions: ingoing into the horizon, and outgoing at infinity.
This turns out to be such a strong requirement that 
the frequency $\omega$ can only take certain discrete values, denoted as $\omega^{(1)}$.
These make up the linear QNM frequency spectrum and, in general, depend on $\ell, m$ and the overtone number
$n$. For a Schwarzschild black hole, the QNM frequency is $m$
independent; not so for a Kerr black hole. Thus, depending on context,
we sometimes use $\omega^{(1)}_{\ell m n}$ and sometimes
$\omega^{(1)}_{\ell n}$ to highlight the mode
dependence of the QNM frequency, though we often suppress these labels
to avoid clutter. The QNM frequency $\omega^{(1)}$ is complex: i.e.
$\omega^{(1)} = \omega_R^{(1)} + i \omega_I^{(1)}$, with
$\omega_I^{(1)} < 0$, signaling decay with time.

It is worth emphasizing that the radial profile of the QNM solution
has an unphysical feature. At $r_* \rightarrow \pm \infty$, 
the QNM mode goes as $e^{-i \omega^{(1)} (t \mp r_*)}$; thus
with $\omega_I^{(1)} < 0$, the QNM mode diverges as $r_* \rightarrow \pm 
\infty$ at a fixed time. Physical perturbations should have no such divergence. 
The best way to think about QNMs is to view them through the lens of
the Green's function whose causal structure ensures such divergence
does not occur \cite{Nollert:1992ifk,Andersson:1996cm,Szpak:2004sf}. 
This will be discussed in detail in the next section.

We will also see how quadratic QNMs arise in the Green's function approach,
but it's not hard to see how they come about at an intuitive level.
Assuming the right hand side of Eq.\ (\ref{QEqn_complex})
is composed of a product of linear modes with time dependence:
$h_{\mu\nu}^{c(1)} \propto
\exp\{-i\omega^{(1)} t\}$ and $\propto
\exp\{-i\omega^{(1)'} t\}$, one can see
the time dependence of $h_{\mu\nu}^{c(2)} \propto \exp\{-i\omega^{(2)}
t\}$ is given by
\begin{equation}\label{QQNM_frequency}
    \omega^{(2)}= \omega^{(1)} + \omega^{(1)'} \quad \text{or} \;
    \quad \omega^{(2)} = \omega^{(1)} - \omega^{(1)'*} \, .
\end{equation}
Thus, for any two linear QNM frequencies $\omega^{(1)}$ and
$\omega^{(1)'}$,
there are two possible quadratic QNM frequencies associated.
Notice that the case with the minus sign can be alternatively thought
of as coming from combining an ordinary linear mode and a mirror mode
(the mirror of an ordinary mode of frequency $\omega$ has frequency
$-\omega^*$ \cite{Berti:2005ys,PhysRevD.103.104048}).
Separating the frequencies into real and imaginary components, we thus have:
\begin{equation}
    \omega_{R}^{(2)} = \omega_{R}^{(1)} \pm  \omega_{R}^{(1)'}; \quad \omega_{I}^{(2)} = \omega_{I}^{(1)} +  \omega_{I}^{(1)'}.
\end{equation}
The quadratic QNM decays with time, since
$\omega_I^{1} , \omega_I^{(1)'} < 0$ implies $\omega_I^{(2)} < 0$, and
in fact decays faster than either of the parent linear QNM modes.
Furthermore, we see that there can be quadratic QNM frequencies that
are purely imaginary (i.e.\ $\omega^{(2)}_R=0$), which will be excited
when a given linear QNM appears in the source with its conjugate
counterpart i.e.\ $\omega^{(2)} = \omega^{(1)} - \omega^{(1)*}=2i\omega_I^{(1)}$. Note that the reasoning used here to obtain the quadratic QNM frequencies is valid for a Schwarzschild or Kerr black hole.

In addition, since the odd and even linear QNM perturbations are isospectral, and all of them can contribute to both odd and even quadratic perturbations, we expect that the same will hold for quadratic modes: the temporal frequency spectrum will be the same for  odd quadratic and  even quadratic QNMs.

From a phenomenological point of view, we emphasize that since the
decay rate of the linear QNMs grows quickly with overtone number $n$,
there will be quadratic QNMs that decay slower than linear
overtones. A particularly relevant QQNM will be the one with harmonic
numbers $(\ell=4, |m|=4)$ since it will be mainly sourced by the
multiplication of two fundamental linear QNMs with ($\ell=2,|m|=2)$\footnote{Note that there are infinite pairs of linear QNM frequencies that will lead quadratic QNMs in the $(4,4)$ harmonic. We have infinite sources coming from the $(2,2,n)$ overtones ($n$ ranging from 0 to $\infty$), as well as infinite combinations of other linear angular harmonics and their overtones.}
(recall Eq.\ (\ref{harmonic_product})), which are the dominant modes
generated from the merger of nearly equal-mass binary black
holes\footnote{In addition, the linear $(\ell=2,|m|=2)$ modes could
  also source quadratic QNMs with $m=0$ and $0 \leq\ell\leq 4$. Such
  quadratic modes would not oscillate in time, but they would decay
  exponentially fast at a rate given by $2\omega^{(1)}_{I\,\,22}$.}.
As an example, for a Schwarzschild black hole of mass $M$, the
$(\ell=2,|m|=2,n=0)$ linear QNM has frequency $GM\omega^{(1)}_{220}=0.374-i0.089$ \cite{Berti:2005ys} and the $(\ell=4,|m|=4,n=0,1)$ linear QNMs have frequencies
$GM\omega^{(1)}_{440}=0.809-i0.094$ and $GM\omega^{(1)}_{441}=0.797-i0.284$.
These frequencies can be compared to that of the QQNM formed by the multiplication of two linear $(2,2,0)$ modes, which gives $GM\omega^{(2)}_{44}=2GM\omega^{(1)}_{ 220}=0.748-i0.178$\footnote{Even though there are infinite quadratic QNM frequencies in (4,4), for simplicity we do not add additional label in the subscript of the quadratic frequency aside from its angular harmonics, and thus implicitly refer to the $(2,2,0)\times(2,2,0)$ quadratic frequency as $\omega^{(2)}_{44}$.} and hence decays slower than the linear $(441)$ mode. 
An analogous behaviour will hold for any spinning black hole, as it can be seen from the general fittings in \cite{Berti_RingData}.
Thus, models of the ($\ell=4,|m|=4$) harmonic in ringdown waveform should include quadratic perturbations. Indeed, \cite{MaganaZertuche:2021syq} analyzed a nearly equal mass non-precessing binary, and found that fitting linear QNMs to NR waveform simulations gives larger residuals of the GW power for ($\ell=4,|m|=4$), compared to other harmonics, suggesting that an improvement in the linear ringdown model is required for $(4,4)$.

The skeptic might argue that the quadratic QNMs could have
very small amplitudes and therefore negligible impact.
However, this does not seem to be the case, as shown in
\cite{London:2014cma}, where analytical fits to (4,4) GWs from NR
simulations were performed and the quadratic (4,4) mode was found to have a comparable amplitude to the linear (4,4) modes.
More generally, nonlinearities are expected to become increasingly
relevant with increasing harmonic numbers (e.g.\ cubic perturbations
could be the leading contribution to the harmonic (6,6), from the multiplication of three linear (2,2,0) QNMs). 

\section{The Green's function approach
}\label{sec:greens}

In this section, we use the Green's function approach to formally
write down the most general first and second-order solutions.
A basic observation is that because the same Green's 
function is used for both, certain features get inherited by
both solutions. 
The Green's function approach has been 
previously used to analyze
linear perturbations \cite{Nollert:1992ifk, Andersson:1996cm, Szpak:2004sf,
  Hui:2019aox}, and second-order ones
\cite{Okuzumi:2008ej}, as well as the BH response to test particles \cite{1977RSPSA.352..381D} and extreme-mass-ratio inspirals (see e.g.\ \cite{Detweiler:2002mi, Poisson:2003nc}).
Much of the discussion in this section is thus a review.
Along the way, we highlight a few 
key lessons that are perhaps not widely appreciated, and work out 
a toy example in great detail to illustrate them.

\subsection{Definitions and setup}
\label{setup}

The Green's function $G$ is defined by:
\begin{eqnarray}
\label{Gdef}
&& \left(-\partial_t^2 + \partial_{r_*}^2 - \hat V\right) G(t, r_*, \theta, \phi
  |\bar t , \bar r_*, \bar \theta, \bar \phi) = \nonumber \\
&& \quad 
\delta(t-\bar t) \delta(r_* - \bar r_*) \delta(\theta - \bar\theta) 
\delta (\phi - \bar\phi) /{\,\rm sin\,}\bar\theta \, ,
\end{eqnarray}
where $\hat V$ is an operator which,
upon acting on (spin-weighted) spherical harmonics, gives rise
to $V_{RW}$ or $V_Z$ (Eqs.\ (\ref{VZ})-(\ref{VRW})).
Time-translation and rotational invariance means it is convenient to
expand the Green's function in terms of Fourier modes 
(in time) and spherical harmonics (in angles):
\begin{align}
\label{Gdecompose}
&G(t,r_*,\theta,\phi| \bar t, \bar r_*, \bar \theta, \bar \phi) =
  \nonumber\\
& \sum_{\ell, m}
  G_{\ell}(t , r_* | \bar t , \bar r_*) \, {}_s Y_{\ell m} (\theta, \phi) {}_s Y_{\ell
  m}^* (\bar \theta,
  \bar \phi) = \nonumber \\
& \sum_{\ell, m}  \int_{\mathbb{C}} {d\omega \over 2\pi} e^{-i \omega (t-\bar t)}
  G_{\omega \ell}(r_* | \bar r_*) {}_s Y_{\ell m} (\theta, \phi) {}_s Y_{\ell
  m}^* (\bar \theta,
  \bar \phi) \, ,
\end{align}
where the integration contour for $\omega$ runs slightly above
the real axis (above all poles of $G_{\omega\ell}$ that end up inside an infinite lower semi-circle; see below), 
such that
$G = 0$ if $t-\bar t < 0$ i.e.\ this is a retarded Green's function.
We have introduced several symbols for the Green's function:
$G$ is the space-time Green's function; 
$G_\ell$ is the 2D Green's function (in radius and time);
$G_{\omega\ell}$ is the radial Green's function.
Substituting this in Eq.\ (\ref{Gdecompose}), we obtain
\begin{eqnarray}
\label{Gdelta}
&& \left(-\partial_t^2 + \partial_{r_*}^2 - V(r_*,\ell) \right) 
G_\ell (t , r_* | \bar t , \bar r_*) = \delta (t-\bar t) \delta (r_* -
  \bar r_*) \, ,\nonumber \\
&& \left(\partial_{r_*}^2 + \omega^2 - V(r_*,\ell) \right) G_{\omega\ell} (r_* |
  \bar r_*) = \delta (r_* - \bar r_*) \, .
\end{eqnarray}
The relevant properties of the spin-weighted spherical
harmonics are their orthonormality and completeness
\cite{1967JMP.....8.2155G}:
\begin{eqnarray}
\label{YellmC}
&& \int {\,\rm sin}\theta \, d\theta d\phi \,
{}_s Y_{\ell m} (\theta, \phi) {}_s Y_{\ell' m'}^* (\theta, \phi)
= \delta_{\ell \ell'} \delta_{m m'} \, , \nonumber \\
&& \sum_{\ell, m} {}_s Y_{\ell m}  (\theta, \phi) {}_s Y_{\ell m}^* (\bar \theta,
  \bar \phi) = \delta(\theta - \bar \theta) \delta(\phi-\bar \phi)/{\,\rm
    sin\,\bar \theta} \, .
\end{eqnarray}
Henceforth, for simplicity, we will set the spin $s=0$, but
it should be kept in mind the entire discussion of this section
can be promoted straightforwardly to any spin $s$ that describes
the fluctuations of interest
\footnote{For instance, the Regge-Wheeler variable (called $Q$ by Regge and Wheeler)
  is defined in terms of the odd part of the metric fluctuation components 
$h_{r\theta}, h_{r\phi}$. Thus it's natural to associate $Q$
with spin $\pm 1$ spherical harmonics. But one could also apply suitable 
spin raising/lowering operators and think of a variable related to $Q$ that is effectively a spin zero quantity,
consistent with the $\ell(\ell+1)$ dependence of $V_{RW}$.
}.
As a comparison, we mention that in the case of a Kerr black hole, $G_\ell$ and $G_{\omega \ell}$ would also depend on the harmonic number $m$, and the spherical harmonics
would be generalized to spheroidal harmonics.

To construct $G_{\omega\ell}$, we need two solutions
$g_{\rm out}$ and $g_{\rm in}$ satisfying
\begin{eqnarray}
\label{g1g2}
\left(\partial_{r_*}^2 + \omega^2-V(r_*,\ell)\right) g_{\rm out, in}(r_*) = 0\, 
\end{eqnarray}
with the desired asymptotic boundary conditions:
$g_{\rm out} (r_*) \rightarrow e^{i\omega r_*}$ as $r_* \rightarrow \infty$
(outgoing at infinity) and
$g_{\rm in}  (r_*) \rightarrow e^{-i\omega r_*}$ as $r_* \rightarrow -\infty$
(ingoing to the horizon), keeping in mind that the potential $V$ vanishes in
both limits. The radial Green's function $G_{\omega\ell}$ can then be constructed as:
\begin{eqnarray}
\label{radialG}
G_{\omega\ell} (r_* | \bar r_*) = {1\over W} g_{\rm out} (r_{*>})g_{\rm in} (r_{*<}),
\end{eqnarray}
where $r_{*>}=\max(r_*,\bar{r}_*)$, $r_{*<}=\min(r_*,\bar{r}_*)$, and $W$ is the Wronskian:
\begin{eqnarray}
W \equiv g_{\rm in} (r_*) \partial_{r_*}
  g_{\rm out} (r_*) - g_{\rm out} (r_*) \partial_{r_*} g_ {\rm
  in}(r_*) \, .
\end{eqnarray}
It is worth noting that $g_{\rm out}$, $g_{\rm in}$ and $W$ depend 
implicitly on $\omega$ and $\ell$, suppressed here to avoid clutter. 
In addition, note that the Wronskian is independent of $r_*$, given the form of Eq.\ (\ref{g1g2}). 

For a general value of $\omega$, the boundary conditions for $g_{\rm
  out}$ and $g_{\rm in}$ cannot be satisfied at the same time and thus
they describe two independent solutions to the homogeneous equation,
and thus $W\not=0$. However, for $\omega$ values that coincide with
the linear QNM spectrum, $g_{\rm out}$ and $g_{\rm in}$ are given by
the same single solution and thus $W=0$. As a consequence, $W$ has first-order \cite{1977RSPSA.352..381D} poles at
the linear QNM frequencies $\omega=\omega^{(1)}_{\ell n}$ (each QNM frequency is labeled by $\ell$ and the overtone $n$; 
for Kerr black holes, there would be $m$ dependence as well).

In general, the exact form of $G_{\omega \ell}$ will depend on the
potential $V$, and for the Zerilli or Regge-Wheeler potentials the
analytical form of $G_{\omega \ell}$ in the full parameter space
$(t,r_*,\bar{t},\bar{r}_*)$ is not known, although its qualitative and
asymptotic features are known
\cite{PhysRevD.34.384,Andersson:1996cm}. 
In particular, after integrating over $\omega$ in 
Eq.\ (\ref{Gdecompose}), the time-domain Green's function can be 
separated into three qualitatively distinct pieces: $G_F$ (flat),
$G_Q$ (QNM), and
$G_B$ (branch cut). 
The piece $G_B$ has to do with the fact that $g_{\rm in}$ and $g_{\rm
  out}$ (and therefore $G_{\omega\ell}$) can have branch cuts in the 
complex $\omega$ plane. Such branch cuts arise from the polynomial
radial decay of the potential (as in the case of $V_Z$ or $V_{RW}$),
and can be understood by back-scattering off it 
\cite{PhysRevLett.74.2414}. We do not have much to say about
this branch-cut contribution $G_B$, other than to note
that it gives rise to signals that tend 
to be subdominant compared to QNM contributions at intermediate times. 

The $G_Q$ piece of the Green's function is associated
with the QNM poles where the Wronksian vanishes.
Recalling the relation between the 2D Green's function
$G_\ell$ 
and the (1D) radial Green's function $G_{\omega \ell}$:
\begin{equation}
G_\ell (t, r_* | \bar t, \bar r_*) = 
\int_{\mathbb{C}}  {d\omega \over 2\pi} e^{-i \omega (t-\bar t)} G_{\omega \ell}(r_* | \bar r_*) \, ,
\end{equation}
the QNM contribution to $G_\ell$ can be written as:
\begin{align}
\label{GQell}
& G_Q {}_\ell (t, r_* | \bar t , \bar r_*)
= \nonumber \\
& \quad \sum_n {-i \over W'_{\ell n}} e^{-i \omega^{(1)}_{\ell n} (t - \bar
  t)} g_\text{out}(r_{*>},\omega^{(1)}_{\ell
  n})g_\text{in}(r_{*<},\omega^{(1)}_{\ell  n})
\Theta
\end{align}
where $\omega_{\ell n}^{(1)}$ is the (linear) QNM frequency, and
$W'_{\ell n} \equiv \partial_\omega W(\omega)$
evaluated at the frequency $\omega^{(1)}_{\ell n}$.
The symbol $\Theta$ schematically represents  causality constraints for  $t, \bar t, r_*, \bar r_*$, which come about depending on 
whether the integration contour in the complex $\omega$ plane can be closed
to include the QNM poles or not; we will see below a more explicit
representation of what this causality constraint entails.

Lastly, $G_{\omega\ell} (r_* | \bar r_*)$ typically has a pole at 
$\omega = 0$ (due not to the Wronskian alone, but its combination
with $g_{\rm out}$ and $g_{\rm in}$ for $V_{RW}$ and $V_Z$). This additional contribution, together with the arcs of the semi-infinite circle of the integration contour is
known as the flat piece of the Green's function $G_F$
(or $G_F {}_\ell$ for the 2D Green's function), and carries information about high-frequency and asymptotically far signals that propagate effectively in free space since they are insensitive to the potential.

To gain more intuition on these different
contributions to the Green's function, it is useful to have explicit
expressions for them. One approach is to display their form
in asymptotic limits; the other is to study a simplified potential
for which closed form analytic expressions are possible.
We show the asymptotic limits in this section, and present
the results of a simplified toy model in Section \ref{app:Vdelta}.

In the large $|r_*|$ limit, where the potential vanishes,
$g_{\rm in}$ and $g_{\rm out}$ behaves as follows
(our discussion follows \cite{, Okuzumi:2008ej}):
\begin{eqnarray}
&& g_{\rm in} \rightarrow e^{-i\omega r_*}  \quad {\rm for} \quad r_* \rightarrow
  -\infty, \nonumber \\
&& g_{\rm in} \rightarrow {\cal A}_{\rm in} e^{-i\omega r_*}  + {\cal B}_{\rm in}
  e^{i\omega r_*} \quad {\rm for} \quad r_* \rightarrow \infty, \\
&& g_{\rm out} \rightarrow {\cal A}_{\rm out} e^{i\omega r_*}  + {\cal B}_{\rm out}
  e^{-i\omega r_*} \quad {\rm for} \quad r_* \rightarrow
  -\infty, \nonumber \\
&& g_{\rm out} \rightarrow e^{i\omega r_*}  \quad {\rm for} \quad r_* \rightarrow
  \infty \, ,
\end{eqnarray}
where ${\cal A}_{\rm in} , {\cal B}_{\rm in} , {\cal A}_{\rm out},
{\cal B}_{\rm out}$ are coefficients that depend on $\omega$ and
$\ell$. The Wronskian $W$ can be computed: $W = 2i\omega {\cal A}_{\rm
  in} = 2i\omega {\cal A}_{\rm out}$. Using these expressions in
(\ref{radialG}) and (\ref{Gdecompose}), it can be shown that
for large $|r_*|$ and $|\bar r_*|$, 
\begin{align}
\label{GFQ}
& G_\ell  (t , r_* | \bar t , \bar r_*)  \sim G_F {}_\ell + G_Q {}_\ell
   \, ,\nonumber \\
& G_F {}_\ell \sim -{1\over 2} \left[ \Theta(t - \bar t - |r_* - \bar r_*|) - 
\Theta(t - \bar t - |r_*| - |\bar r_*|) \right] \, ,\nonumber \\
& G_Q {}_\ell \sim \sum_n {-i f_{\ell n}\over W'_{\ell n}}
e^{-i \omega_{\ell n}^{(1)} (t - \bar t - |r_*| - |\bar r_*|)} 
\Theta(t - \bar t - |r_*| - |\bar r_*|),
\end{align}
where $\Theta(x)$ is the step function (unity if $x > 0$, zero
otherwise). 
The factor $f_{\ell n}$ is an order unity function of $r_*$, $\bar r_*$ and
$\omega^{(1)}_{\ell n}$
\footnote{
More precisely, $f_{\ell n} = 1$ if $r_*$ and $\bar r_*$ have opposite signs, 
$f_{\ell n} = \mathcal{B}_{\rm in}$ evaluated at $\omega^{(1)}_{\ell n}$ if both $r_*$ and $\bar r_*$
are positive, and $f_{\ell n} = \mathcal{B}_{\rm out}$ evaluated at $\omega^{(1)}_{\ell n}$ if
both $r_*$ and $\bar r_*$ are negative. 
The derivation goes roughly as follows: for instance,
for $r_* > 0$ and $\bar r_* < 0$ (and both large in magnitude), 
$g_{\rm in} (r_*) g_{\rm out} (\bar r_*) = e^{i\omega (r_* - \bar
  r_*)}$ giving rise to $G_Q {}_\ell$ with $f_{\ell n} = 1$.
For $r_* > \bar r_* > 0$, $g_{\rm in} (r_*) g_{\rm out} (\bar r_*) = 
{\cal A}_{\rm in} ( e^{i\omega (r_* - \bar r_*)} - e^{i\omega (r_* +
  \bar r_*)} ) + ({\cal A}_{\rm in} + {\cal B}_{\rm in} )e^{i\omega (r_* +
  \bar r_*)}$: the first term gives $G_F {}_\ell$, and the second term
gives $G_Q {}_\ell$ with the appropriate $f_{\ell n}$, keeping in mind
$W = 2 i\omega {\cal A}_{\rm in}$, and
${\cal A}_{\rm in}$ vanishes at the linear QNM frequencies.
}.
It is worth stressing the limitation of 
(\ref{GFQ}): it ignores the branch-cut contribution and holds only
for large $|r_*|$ and $|\bar r_*|$, which is not useful for realistic calculations but it nevertheless helps illustrates the main properties of $G_{\ell}$.

\begin{figure}[h!]
\centering
\includegraphics[width = 0.40\textwidth]{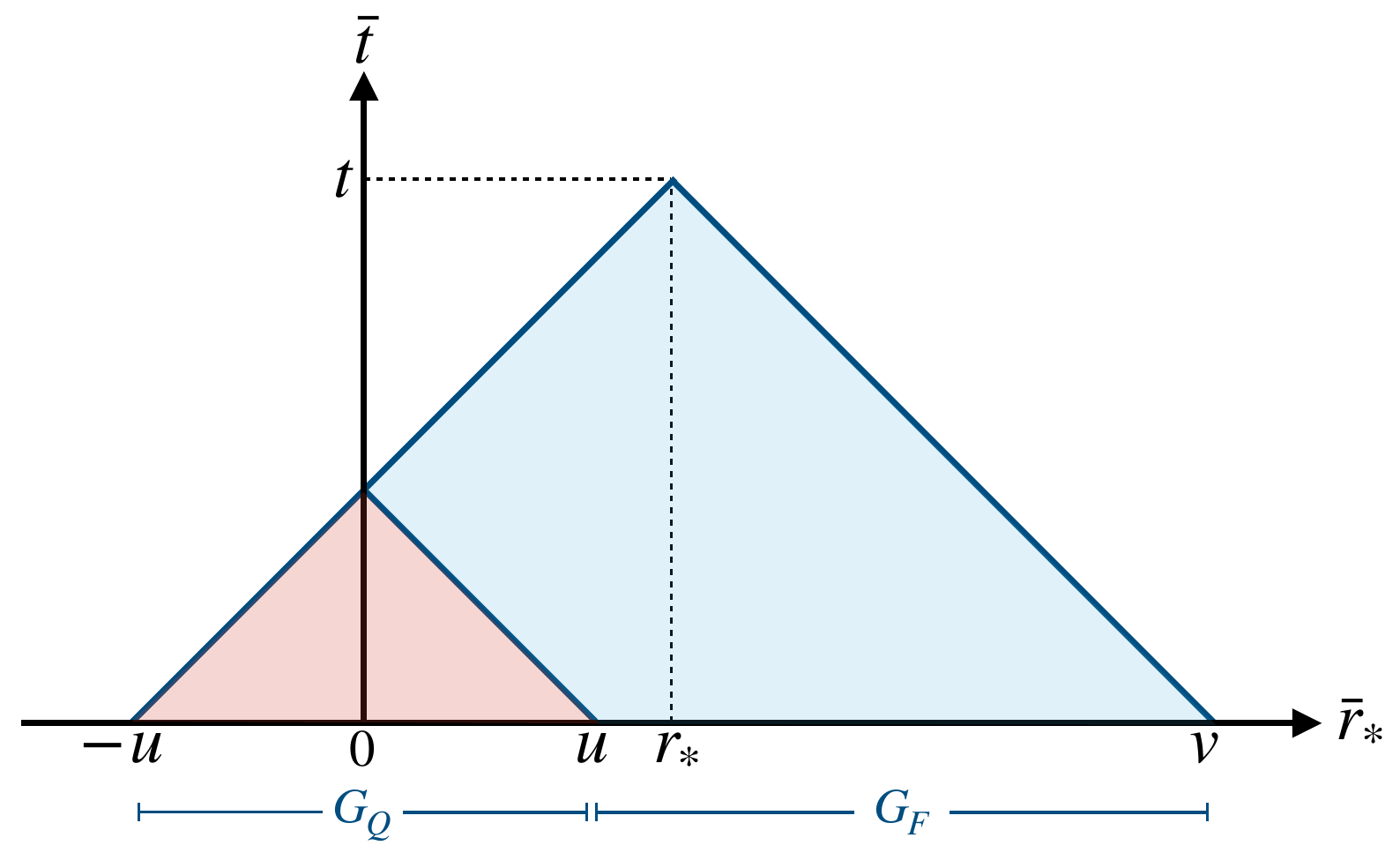}
 \caption{Support of asymptotic $G_{F\ell}$ (shaded blue) and $G_{Q \ell}$ (shaded red) for a given point $(t,r_*)$.  Here, $u=t-r_*$ and $v=t+r_*$, and the potential peak is around $\bar{r}_*=0$. In general, the boundaries of these regions are expected to be fuzzy but this figure schematically illustrates the role of causality constraints. Horizontal blue lines indicate maximum size of spatial region causally connected to $(t,r_*)$ through $G_{Q\ell}$ and $G_{F\ell}$.}
 \label{Gsupportv2}
\end{figure}

The step functions in the above expressions represent
non-trivial causality constraints coming from how the contour
in the $\omega$ integral closes. In particular, the step function for $G_Q {}_\ell$ tells us
the QNM piece of the Green's function
does not vanish only if the point $(\bar t , \bar r_*)$ is
causally connected to $(t, r_*)$ via the potential, as illustrated in Fig.\ \ref{Gsupportv2}.

In the asymptotic form given for $G_Q {}_\ell$, the potential can be 
roughly thought
of as being located at small tortoise radii (i.e. in the vicinity of the origin).
In reality of course, neither $V_{RW}$ nor $V_Z$ is well localized
(though they do peak at a small tortoise radius);
the step function in $G_Q {}_\ell$ presents what is more akin to
a bird's-eye view of how it behaves. In particular, the step function
tells us $G_Q {}_\ell$  vanishes when $|\bar r_*|$ is too big, 
i.e. if $|\bar r_*|$ veers too far from where the
potential peaks (and the larger $t - \bar t - |r_*|$ is, the further
$|\bar r_*|$ can veer). 
The step functions in $G_F {}_\ell$, on the other hand, combine
to constrain $\bar r_*$ to be within the past light-cone of $t, r_*$,
but away from the regions where the potential is non-negligible (see Fig.\ \ref{Gsupportv2}).

With our bird's-eye view of the Green's function (\ref{GFQ}), 
i.e. valid only at large tortoise radius (or absolute value thereof), 
let us introduce one small improvement. The expressions
given in (\ref{GFQ}) privileges the origin, as if the potential is
located there. In practice, if there's a privileged position, it ought
to be the location of the top of the potential. For instance, 
in the WKB approach to computing the linear QNM spectrum, it is
the derivatives of the potential at the top that determines the QNM
frequencies. Henceforth, when we use (\ref{GFQ}), we will replace
$r_* \rightarrow r_* - \hat r_*$ and $\bar r_* \rightarrow \bar r_* -
\hat r_*$, with $\hat r_*$ representing the location of the 
potential peak. In other words, within the large radius approximation
that led to (\ref{GFQ}), there is effectively no difference between
$r_*$ and $r_* - \hat r_*$, or between $\bar r_*$ and
$\bar r_* - \hat r_*$, as long as $\hat r_*$ is small, which it is for $V_{RW}$ and $V_Z$.

\subsection{First-order perturbations}
\label{linearEvolve}

We first review how the Green's function is used to evolve the first-order
perturbations. Consider a first-order perturbation:
\begin{equation}
\Psi^{(1)} (t, r_* , \theta, \phi) = \sum_{\ell, m} \Psi^{(1)}_{\ell
  m} (t, r_*) Y_{\ell m} (\theta, \phi) \, , 
\end{equation}
satisfying 
\begin{equation}
(-\partial_t^2 + \partial_{r_*}^2 - V) \Psi^{(1)}_{\ell
  m} (t, r_*) = 0 \, .
\end{equation}
Recall again all expressions here can be promoted to spherical harmonics
of any spin-weight.
Let us define the initial conditions to be:
\begin{eqnarray}
\psi_0 (r_*) \equiv \Psi^{(1)}_{\ell m} (0, r_*) , \quad
\dot\psi_0 (r_*) \equiv \partial_t \Psi^{(1)}_{\ell m} (t, r_*)
  |_{t=0} \;,
\end{eqnarray}
where we have suppressed the $\ell, m$ dependence of $\psi_0$ and
$\dot\psi_0$ to avoid clutter. Henceforth, $t=0$ is adopted as the initial time.

The Green's function can be used to evolve the linear perturbation
forward as: 
\begin{align}
    \Psi^{(1)}_{\ell m} (t,r_*) =\int d\bar{r}_* \left[ \partial_{\bar{t}}G_\ell
      |_{\bar{t}=0}\psi_{0}(\bar{r}_*)- G_\ell |_{\bar{t}=0}\dot{\psi}_{0}(\bar{r}_*)\right],\label{Phi1_Gral} 
\end{align}
where $G_\ell(t, r_* | \bar t , \bar r_*)$ represents the 2D Green's
function defined in Eq.\ (\ref{Gdecompose}). Its retarded nature means it
vanishes unless $t > \bar t$. The derivation of this standard result
can be found in e.g. \cite{1962clel.book.....J,Szpak:2004sf,Hui:2019aqm}.

Making use of (\ref{Gdecompose}), we can also write this as: 
\begin{align}
   \Psi^{(1)}_{\ell m}(t,r_*)=  & \int d\bar{r}_* \int
     \frac{d\omega}{2\pi}e^{-i\omega t}G_{\omega \ell}(r_*|\bar{r}_*) 
\nonumber \\ & \left[i\omega \psi_{0}(\bar{r}_*) - \dot{\psi}_{0}(\bar{r}_*)\right] .\label{Psi1_G}
\end{align}

Making use of the flat/QNM/branch-cut split of the Green's function, we can split the linear solution as:
\begin{equation}
\Psi^{(1)}_{\ell m} (t , r_*) = \Psi^{(1)}_F {}_{\ell m} (t , r_*) 
+ \Psi^{(1)}_Q {}_{\ell m} (t , r_*) +  \Psi^{(1)}_B {}_{\ell m} (t , r_*) \, .\label{Psi1_Green_split}
\end{equation}
From the asymptotic solutions in (\ref{GFQ}), we can see that $\Psi^{(1)}_F$ gives rise to waves traveling to the left
(horizon) or to the right (infinity) that reflect the initial
conditions. We will work this out in detail in
Section \ref{app:Vdelta}, for a toy example where
$G_F {}_\ell$ given in (\ref{GFQ}) is exact. In addition, it is known that $\Psi^{(1)}_B$ leads to polynomial tails due to the long-range polynomial decay of $V_Z$ and $V_{RW}$ \cite{PhysRevD.34.384, Andersson:1996cm, PhysRevLett.74.2414}.
It is the QNM piece of the Green's function $G_Q {}_\ell$ that gives rise to a signal oscillating at the QNM frequencies. 

The QNM part of the linear perturbation is:
\begin{align}
\label{Psi1_Q}
   \Psi^{(1)}_Q {}_{\ell m} (t, r_*) = & \int d\bar r_* \sum_n {-i
                                         \over {W'_{\ell  n}}}
                                         e^{-i\omega^{(1)}_{\ell  n}
                                         t}
                                         g_\text{out}(r_{*>},\omega^{(1)}_{\ell
                                         n}) \nonumber \\ & \quad
                                                            g_\text{in}(r_{*<},\omega^{(1)}_{\ell n})
    \left[i\omega^{(1)}_{\ell  n} \psi_{0}(\bar{r}_*) -
      \dot{\psi}_{0}(\bar{r}_*)\right] \Theta
\nonumber \\ \sim  & \int d\bar r_* \sum_n {-i
     f_{\ell n} \over W'_{\ell n}} e^{-i \omega_{\ell n}^{(1)} (t - |r_* -
                     \hat r_*| - |\bar r_* - \hat r_*|)}
     \nonumber \\
& \,\, \left( i
     \omega_{\ell n}^{(1)} \psi_0(\bar r_*) - \dot\psi_0(\bar r_*)
  \right) \nonumber \\
& \,\,\, \, \Theta(t
     - |r_* - \hat r_*| - |\bar r_* - \hat r_*|) \, .
\end{align}
The first equality follows from (\ref{GQell})\footnote{Due to Eq.\ (\ref{Phi1_Gral}), we expect additional terms coming from taking the derivative of $G_Q$ and this derivative acting onto the $\Theta$ function. For simplicity we have omitted this extra terms here but they will be shown explicitly in the toy example of Subsec.\ \ref{app:Vdelta}.}, whereas in the second equality we have used the asymptotic expression and abused (\ref{GFQ}) a bit: 
(\ref{GFQ}) is meant for large $|\bar r_*|$ 
(and $| r_* |$), while the above integral ranges over all values of
$\bar r_*$. Nonetheless, a few important points stand:
(1) The first-order perturbation acquires oscillatory behavior at the
QNM frequencies, regardless of details of the initial conditions
(codified by $\psi_0$ and $\dot\psi_0$).
(2) The QNM part of the Green's function vanishes if $\bar r_*$ 
is too large, due to the causality constraint signified
by the step function. Thus, the integral over $\bar r_*$ is limited
to regions around the peak of the potential (with a range determined
by $t - |r_*-\hat r_*|$) \cite{Andersson:1996cm, Szpak:2004sf, Hui:2019aox}. 
(3) Because the range of $\bar r_*$ that contributes
to the integral is time-dependent, the QNM oscillations in general
have time-dependent amplitudes---this is true even within linear
perturbation theory. Thus, in analyzing numerical/observational
ringdown data, the time-dependent nature of the amplitudes of QNM oscillations
should not be interpreted, on its own, as evidence for the break down of linear
perturbation theory. This raises the interesting question of what precise
model to use when fitting numerical or detected signals with QNMs, especially close to the merger time. We will illustrate this amplitude variation in a toy example in Section \ref{app:Vdelta} .

Henceforth, we approximate the QNM part of the linear perturbation as:
\begin{equation}
\label{Psi1QA}
\Psi^{(1)}_Q {}_{\ell m} (t, r_*) \sim \sum_n A (t, r_*) e^{-i
  \omega^{(1)}_{\ell n} (t - |r_*-\hat r_*|)} \Theta(t - |r_* - \hat r_*|) \, ,
\end{equation}
where $A(t, r_*)$ represents the result of the integral over $\bar
r_*$. If the initial conditions $\psi_0$, $\dot \psi_0$ were
sufficiently localized around the peak of the potential, then $A$ would be 
time-independent after some amount of time (such that $t - |r_* - \hat
r_*|$
covers the entire range of $\bar r_* - \hat r_*$ over which the initial
conditions were non-vanishing); otherwise, $A$ may depend on
time. Note we have suppressed the $\ell, n$ and $\omega^{(1)}_{\ell n}$ dependence of $A$ to simplify notation.

The remaining step function $\Theta(t - |r_* - \hat r_*|)$ in
Eq.\ (\ref{Psi1QA}) is important: it tells us that if 
$t < |r_* - \hat r_*|$ (i.e the location of interest is too far away
relative to the time of interest), there's no value of $\bar r_*$
that would satisfy the causality condition for producing QNMs,
and so the integral (\ref{Psi1_Q}) vanishes. In other words,
the linear QNM oscillations are visible only to someone at
a location $r_*$ and time $t$ that is causally connected to
the bulk of the potential (represented by its peak).
The combined presence of $A(t, r_*)$ and $\Theta(t - |r_* - \hat
r_*)$ tells us the actual theoretical prediction for the
observable linear perturbations does {\it not} have the
precise classic form of a QNM ${\rm exp\,} [{-i
  \omega^{(1)}_{\ell n} (t - |r_*-\hat r_*|)}]$, but is instead
modulated. In particular, at a fixed time $t$, 
the linear perturbations do not exponentially diverge at large radius,
despite the frequency having a negative imaginary part (see further detailed discussions of causality in \cite{Szpak:2004sf}).

\subsection{Second-order
perturbations}
\label{secondGenerate}

Consider next the generalization of Eq.\ (\ref{Gdef}) to an arbitrary
source:
\begin{eqnarray}
\left( -\partial_t^2 + \partial_{r_*}^2 - \hat V \right) \Psi^{(2)} (t, r_*,
  \theta, \phi) = S^{(2)} (t, r_*,
  \theta, \phi) .\label{Psi_Eqn_Greens}
\end{eqnarray}
We are interested in $S^{(2)}$ consisting of quadratic
combinations of first-order perturbations, sourcing the second-order
perturbations $\Psi^{(2)}$. 
The solution to this equation generally contains both homogeneous and particular pieces. The homogeneous solution will be determined by  initial conditions on $\Psi^{(2)}$, and it will behave exactly as the linear QNMs $\Psi^{(1)}$. For this reason, we will assume that, if perturbation theory works, all the initial conditions will be attributed to $\Psi^{(1)}$, and $\Psi^{(2)}$ will vanish initially. Let us then focus on the particular solution of $\Psi^{(2)}$ due to the source, which can be written in terms of the Green's function $G(t,r_*,\theta,\phi| \bar t, \bar r_*, \bar \theta, \bar \phi)$ as follows:
\begin{align}
\label{PsiG}
\Psi^{(2)} (t, r_*, \theta, \phi) &= \int d\bar t d\bar r_* d\bar
  \theta d\bar\phi {\,\rm
  sin\,}\bar \theta \, G(t,r_*,\theta,\phi| \bar t, \bar r_*, \bar \theta, \bar \phi)\nonumber\\
  &\times  S^{(2)}(\bar t, \bar r_*, \bar \theta, \bar \phi) \, ,
\end{align}
where the Green's function can be decomposed in frequency-harmonic
space following (\ref{Gdecompose}).

The source $S^{(2)}$ is composed of many quadratic combinations
of the linear perturbations. We are particularly interested in
linear perturbations that contain the (linear) QNM oscillations. 
Consider thus the following illustrative source, from
``squaring'' (\ref{Psi1QA}):
\begin{align}
& S^{(2)} (t, r_*, \theta, \phi)  = \left(A_1 e^{-i \omega_1 (t - |r_* - \hat
  r_*|)} Y_{\ell_1 m_1} (\theta, \phi) + {\,\rm c.c.} \right)\nonumber \\
& \quad \times \left(A_{1'} e^{-i \omega_1' (t -
  |r_* - \hat r_*|)} Y_{\ell_1' m_1'} (\theta, \phi) + {\, \rm c. c.} \right) \Theta (t - |r_* - \hat
r_*|) \, .
\end{align}
Here we assume the source is real, but a complex source
can be dealt with following Eq.\ (\ref{QEqn_complex}).
We use $A_1, \omega_1, \ell_1, m_1$ and
$A_{1'}, \omega_1', \ell_1', m_1'$ to denote
properties of the two linear QNMs \footnote{As discussed in Section \ref{sec:QNM_Eqn}, it is desirable to have a source that falls off at infinity and at the horizon. One can think of these additional fall-off factors as absorbed into the definition of the amplitudes $A_1$ and $A_{1'}$. See 
Section \ref{app:Vdelta} for a concrete example.}.

Using the Clebsh-Gordan coefficients (\ref{harmonic_product}), the
source can be rewritten as:
\begin{align}
    &S^{(2)}(t, r_*,  \theta,  \phi) =
      \sum_{\ell=|\ell_1-\ell_1'|}^{\ell=\ell_1+\ell_1'}  \left[e^{-i
      \omega_{+}(t-|r_* - \hat r_*|)}  {A}_{1}  {A}_{1'}  c_{\ell m_+} Y_{\ell m_+}( \theta, \phi)\right. \nonumber\\
    &\left. +e^{-i \omega_{-}(t-|r_* - \hat r_*|)}  {A}_{1}  {A}_{1'}^{*} c_{\ell m_{-}}(-1)^{m_1'} Y_{\ell m_{-}}( \theta, \phi)+ {\rm c. c.} \right]\nonumber\\
    & \times\Theta(t-|r_*-\hat{r}_*|), \label{S2_gral}
\end{align}
where $c_{\ell m}$ are angular-mixing coefficients that appear on the right-hand side of Eq.\ (\ref{harmonic_product}), and we have defined $m_{\pm}=m_1\pm m_1'$ as well as the frequencies $\omega_+ \equiv \omega_1 + \omega_1'$ 
and $\omega_- \equiv \omega_1 - \omega_1'^*$ that were discussed in 
Section \ref{sec:angle_time}
\footnote{
Here, we use $\omega_1, \ell_1, m_1$ and $\omega_1', \ell_1', m_1'$ to
denote properties of the two linear QNMs, and $\omega_\pm, \ell,
m_\pm$ for the corresponding second-order QNMs.
Elsewhere in the paper, we use $\omega, \ell, m$
and $\omega', \ell', m'$ to denote properties of the two linear QNM
modes and $\omega_2, \ell_2, m_2$ for the corresponding second QNMs.
Also, occasionally, to emphasize
that $\omega, \omega'$ refer to frequencies of linear modes, we use
$\omega^{(1)}$ and $\omega^{(1)'}$. And likewise 
$\omega^{(2)}$ for the frequency of the quadratic mode. }.

Next, we calculate the second-order solution by substituting the source (\ref{S2_gral}) into Eq.\ (\ref{PsiG}). We first perform the angular integral as well as $\bar{t}$ integral from $|\bar{r}_*-\hat{r}_*|$ to infinity, assuming that $\omega$ is slightly above the real axis (i.e.\ with positive imaginary part):
\begin{align}
\Psi^{(2)} =& \sum_{\ell=|\ell_1-\ell_1'|}^{\ell=\ell_1+\ell_1'}  \left[c_{\ell m_+} Y_{\ell m_+} (\theta, \phi) I_{\ell +}(t, r_*) \right.\nonumber\\
& \left.+ \, c_{\ell m_-} (-1)^{m_1'}Y_{\ell m_-} (\theta, \phi) I_{\ell -}(t, r_*) + {\,\rm c. c.}\right], 
\end{align}
where 
\begin{align}
  I_{\ell +} &= -i \int d\bar{r}_*{A}_{1} (\bar r_*) {A}_{1'} (\bar
                 r_*) \int\frac{ d\omega }{2\pi} \frac{G_{\omega
                 \ell}(r_*|\bar{r}_*)}{(\omega-\omega_+)} e^{-i\omega
                 (t-|\bar{r}_*-\hat{r}_*|)},\label{Iplus} \\
  I_{\ell -} &= -i \int d\bar{r}_*{A}_{1} (\bar r_*) {A}^*_{1'} (\bar
               r_*) \int\frac{ d\omega }{2\pi} \frac{G_{\omega
               \ell}(r_*|\bar{r}_*)}{(\omega-\omega_-)} 
e^{-i\omega
              (t-|\bar{r}_*-\hat{r}_*|)} .
\label{Iminus}
\end{align}
In performing the integral over $\bar t$, which gives us the factor
of $\omega - \omega_\pm$ in the denominator, we have assumed $A_1$ and
$A_{1'}$ are independent of time. As discussed earlier (below
Eq.\ (\ref{Psi1QA})), this is not
true in general, but they might vary slowly enough compared to the time
scale set by $\omega_\pm$ or asymptote to constant values. 
Here we see that the integrand in $\omega$ now has poles at the linear
QNM frequencies coming from the $G_{\omega\ell}$, as well as poles at
the frequencies $\omega_{\pm}$ of the quadratic source. While in
general we expect the quadratic and linear frequencies to be
different, a previous analysis shows that there may be enhancements of
the excited amplitudes when the source has a frequency (given by
$\omega_{\pm}$ in our setup) close to the natural frequencies of the
black hole (given by $\omega^{(1)}$), in analogy to resonance
\cite{1977RSPSA.352..381D}. To what extent resonance is important for quadratic
QNMs is a subject we will return to in the future.

If we were to perform the integrals in Eq.\
(\ref{Iplus})-(\ref{Iminus}), we again expect three distinct
contributions to be present in the second-order solution, coming from
$G_F$, $G_Q$ and $G_B$. The solution coming from $G_F$ has been
studied asymptotically in \cite{Okuzumi:2008ej} (using
expressions in 
Eq.\ (\ref{GFQ})), where it was found that $\Psi^{(2)}_F$ will
have QNM ringing solutions at the quadratic frequencies $\omega_{\pm}$
as well as polynomial tails when the quadratic source has a long-range
polynomial decay. Intuitively, since $G_F$ approximates to a flat
space propagator, it is expected to induce solutions with an analogous
functional form as the quadratic source. Mathematically, the fact that
quasi-normal modes with  $\omega_{\pm}$ appear from $G_F$ is expected
from Eqs.\ (\ref{Iplus})-(\ref{Iminus}) since any term in $G_{\omega
  \ell}$---in particular, those that generate $G_F$---now has extra
poles at $\omega_{\pm}$ that need to be taken into account in the
frequency integral.

In addition, we can analyze the second-order solution related to
$G_Q$. For this, we include both the poles associated with the
vanishing of the Wronskian (located at the linear QNM frequencies),
and the new pole associated with the frequencies $\omega_\pm$.
In that case, from the frequency integral we expect to obtain terms like:
 \begin{align}
    & I_{\ell +} \supset  - \int d\bar{r}_*{A}_{1} (\bar r_*) {
  A}_{1'} (\bar r_*)  \Bigg[ G_{\omega_+
      \ell}(r_*|\bar{r}_*)e^{-i\omega_+ (t - |r_* - \hat r_*|)}\nonumber\\
 & \left. + \sum_{n} \frac{g_\text{out}(r_{*>},\omega^{(1)}_{\ell
   n})g_\text{in}(r_{*<},\omega^{(1)}_{\ell n})}{W'_{\ell
   n}(\omega^{(1)}_{\ell n}-\omega_+)}e^{-i\omega^{(1)}_{\ell n}(t - |\bar{r}_*-\hat{r}_*|)}\right], \label{Iplus}\\
     & I_{\ell -} \supset  - \int d\bar{r}_*{A}_{1} (\bar r_*) {
  A}^*_{1'} (\bar r_*)  \Bigg[ G_{\omega_- \ell}(r_*|\bar{r}_*)e^{-i\omega_- (t- |r_* - \hat r_*|)}\nonumber\\
 & \left. + \sum_{n} \frac{g_\text{out}(r_{*>},\omega^{(1)}_{\ell
   n})g_\text{in}(r_{*<},\omega^{(1)}_{\ell n})}{W'_{\ell
   n}(\omega^{(1)}_{\ell n}-\omega_-)}e^{-i\omega^{(1)}_{\ell n}
   (t - |\bar{r}_*-\hat{r}_*|)} \right].\label{Iminus}
\end{align}
It is worth noting that these expressions can typically be simplified if we are interested in $\Psi^{(2)}$ for asymptotically far observers, as in that case $g_{\rm out} (r_*) \approx  e^{i\omega^{(1)}_{\ell n} r_*}$ and $g_{\rm in} (r_{*<}) = g_{\rm in} (\bar r_*)$, assuming $A_1$ and $A_{1'}$ vanish at sufficiently large $\bar r_*$.

From Eqs.\ (\ref{Iplus})-(\ref{Iminus}) we first see that the second-order solution from $G_Q$,
$\Psi^{(2)}_Q$, will generally contain QNMs at the linear frequencies
$\omega^{(1)}_{\ell n}$. This result shows that the linear QNM amplitudes receive non-linear corrections, which agrees with previous numerical results \cite{Okuzumi:2008ej,Ripley:2020xby} that have observed quadratic excitations evolving at the linear frequencies. In addition, here we find that $G_Q$ also leads to further terms that evolve at the quadratic frequencies $\omega_{\pm}$ (in contrast to what was suggested in \cite{Okuzumi:2008ej}). An important difference is that a given quadratic frequency $\omega_\pm$ is only sourced by one specific pair of linear QNMs in the quadratic source, whereas a given linear frequency $\omega^{(1)}_{\ell n}$ is expected to be sourced by an infinite number of pairs of linear QNMs in the  quadratic source. This happens because $\omega^{(1)}_{\ell n}$ are characteristic frequencies of the Green's function (and not a sole property of linear theory) and thus any source, regardless of its shape, is expected to excite these characteristic frequencies. 

In the next subsection, we will use a toy model to qualitatively confirm these results and show that $\Psi^{(2)}$ will indeed contain both quasi-normal modes at the linear frequencies $\omega^{(1)}$ (from $G_Q$) as well as quadratic frequencies at $\omega_{\pm}$ (from $G_F$ and $G_Q$). 

Finally, from $G_B$, we expect the second-order solution
$\Psi^{(2)}_B$ to have polynomial tails (in analogy to the first-order
solution) as well as some exponentials in time with $\omega_{-}$
frequencies. This is because the solution associated to $G_B$ is
obtained by integrating over a branch-cut line for purely negative
imaginary values of $\omega$, and sometimes $\omega_{-}$ can lie along
that line (when the two linear QNMs in the quadratic source are the
same and one of them is conjugated). Thus, the integrand that gives
$\Psi^{(2)}_B$ will have $\omega_{-}$ poles along the branch cut that
need to be taken into account, by deforming the integration contour
around these poles in the complex plane \footnote{Another intuitive
  way of understanding that we should have QNM solutions with purely
  imaginary $\omega_{-}$ frequencies from $G_B$ is to note that the
  choice of the branch cut location is convention dependent, and we could have chosen it not to be along the purely negative imaginary axis, in which case the poles $\omega_{-}$ would have become part of the residue integral and behaved as any other QNM term found in $\Psi^{(2)}_Q$.}. Note however, that these $\omega_{-}$ modes will describe purely exponentially decaying modes that do not oscillate in time, and can be interpreted as transitory memory effects.

We emphasize that these qualitative results can be straightforwardly generalized to $j$-th order perturbations since we expect to have the same starting equation (\ref{Psi_Eqn_Greens}) but with a source composed of various multiplications of perturbations of order lower than $j$. In particular, we expect to excite oscillatory modes with frequencies $\omega^{(j)}$ that are $j$ additions and/or subtractions of linear QNM frequencies and their conjugates, as well as polynomial tails, and oscillatory QNMs with linear frequencies $\omega^{(1)}$. Therefore, we expect the linear QNM spectrum to receive amplitude corrections at all non-linear orders.

\subsection{Example: delta function potential}\label{app:Vdelta}
In this section we consider a simple model where we can calculate analytically the first and second-order solutions using the Green's function approach. Let us consider the following $1+1$ starting equation of motion:
\begin{equation}
    \left(-\partial_t^2+\partial_x^2 - V_0\delta (x)\right)\Psi^{(1)}(t,x)=0,\label{Eq_Delta}
\end{equation}
where $x$ is analogous to the tortoise coordinate, and ranges between $-\infty$ to $+\infty$. Here, we also introduce a potential parameter $V_0>0$ so that the potential is positive and located at $x=0$. This is a toy model in which  $x=0$ is analogous to the location where the RW and Zerilli potentials peak. This potential was studied in e.g. \cite{Hui:2019aox}. 

The retarded Green's function for Eq.\ (\ref{Eq_Delta}) is given by \cite{Hui:2019aox}:
\begin{align}
    &G(t,x|\bar{t},\bar{x}) =  G_F (t,x|\bar{t},\bar{x}) +  G_Q (t,x|\bar{t},\bar{x}),
\end{align}
where
\begin{align}
    & G_F (t,x|\bar{t},\bar{x}) = -\frac{1}{2}\left[\Theta(t-\bar{t}-|x-\bar{x}|)-\Theta( t  - \bar{t} - |x| - |\bar{x}|)\right], \label{GF_Delta}\\
    & G_Q (t,x|\bar{t},\bar{x}) = -\frac{1}{2}e^{-\frac{V_0}{2}(t-\bar{t}-|x|-|\bar{x}|)}  \Theta(t-\bar{t}-|x|-|\bar{x}|),\label{GQ_Delta}
\end{align}
where $G_F$ does not depend on the potential $V_0$ and thus it propagates signals to the observer through flat space, whereas $G_Q$ depends on the only linear QNM frequency present in this example $\omega^{(1)}=-iV_0/2$ (which happens to be purely imaginary) and propagates signals that get transmitted or reflected by the potential. 
Comparing the above with (\ref{GFQ}) is instructive: what was approximately true (in asymptotic limits) is now exactly true for all $x$ and $\bar x$.

Given some initial conditions $\psi_0(x)$ and $\dot{\psi}_0(x)$, the total linear solution will contain two pieces, coming from $G_F$ and $G_Q$. In the former case, we replace Eq.\ (\ref{GF_Delta}) into Eq.\ (\ref{Phi1_Gral}) (without angular dependence) and obtain:
\begin{align}
    \Psi^{(1)}_F(t,x)&=\frac{1}{2}\left(\psi_0(-u)+\psi_0(v)\right)+\frac{1}{2}\int_{-u}^v d\bar{x} \;\dot{\psi}_0(\bar{x})\nonumber\\
    &-\frac{1}{2}\Theta(t-|x|)\left(\psi_0(|x|-t)+\psi_0(t-|x|)\right)\nonumber\\
    & - \frac{1}{2}\Theta(t-|x|)\int_{|x|-t}^{t-|x|} d\bar{x} \;\dot{\psi}_0(\bar{x}),\label{PsiF1_Delta}
\end{align}
where we have defined $u=t-x$ and $v=t+x$.
The first line describes free propagating waves in any direction, that would always be present, even in the absence of a potential. The second and third lines describe the region that is causally connected to the potential at $x=0$ and that hence should not describe completely free waves and this is why it has opposite signs to the free solution. This happens because $G_F$ contains information about the existence of the potential (through the second step function in Eq.\ (\ref{GF_Delta})) but not to its properties. Therefore, all the free waves generated by $G_F$ vanish at $x=0$. These waves have an analogous behaviour to those in a string with a fixed end at a wall.
As a consequence, from (\ref{PsiF1_Delta}) we see that  the solution $\Psi^{(1)}_F$ for $x<0$ only depends on the value of the initial conditions at $x<0$, and the same holds for $x<0$. 
In an analogy with a Schwarzschild black hole, this means that the free waves traveling close to the horizon only depend on what was the initial condition close to the horizon, and that asymptotically far observers are only sensitive to the initial conditions to the right of the potential. Therefore, if the initial conditions happen to be large for $x<0$ and small for $x>0$ (as one may expect in the case of isolated binary black hole mergers), then asymptotically far observers will detect a small signal $\Psi^{(1)}_F$ at any time. 
This result emphasizes the need for distinguishing and modelling differently asymptotically far GWs versus the entire GW radial profile.

Next, we calculate the linear solution coming from $G_Q$.  Substituting Eq.\ (\ref{GQ_Delta}) into Eq.\ (\ref{Phi1_Gral}) we obtain (analogous to (\ref{Psi1_Q}) whose approximation is now exact):
\begin{align}
    \Psi^{(1)}_Q(t,x)&= A(t,x) e^{-\frac{V_0}{2}(t-|x|)}\Theta(t-|x|)\label{Psi1_GralAmplitude}\\
    & + \frac{1}{2}\left[\psi_0(t-|x|) +\psi_0(|x|-t) \right]\Theta(t-|x|)\,\label{Psi1_QFree} ;\\
    A(t,x) & =\frac{1}{2} \int_{|x|-t}^{t-|x|}d\bar{x}e^{\frac{V_0}{2}|\bar{x}|}\left[\dot{\psi}_0(\bar{x})-\psi_0(\bar{x})\frac{V_0}{2}\right]\label{A_Integral}
\end{align}
From here we see that $\Psi_Q^{(1)}$ has two pieces. On the one hand, (\ref{Psi1_GralAmplitude}) looks analogous to the usual QNM models used in the literature, that contains an exponential with the linear QNM frequency $\omega^{(1)}=-iV_0/2$ and the radiation is outgoing at spatial infinity and ingoing at the horizon. On the other hand, (\ref{Psi1_QFree}) contains free travelling waves in the region causally connected to the potential peak, and cancels out the second line of Eq.\ (\ref{PsiF1_Delta}) in order to recover free-space waves when $V_0=0$. From now on, we then continue focusing just on (\ref{Psi1_GralAmplitude}).

Importantly, the solution  (\ref{Psi1_GralAmplitude}) includes a causality condition imposed by the theta function, which avoids divergences in the limit of large $|x|$. In addition, notice that this linear solution describes a wave that always propagates away from the potential, which is what happens in the eikonal limit for linear QNMs of a Schwarzschild black hole around the light ring, as seen in \cite{1985ApJ...291L..33S} and confirmed in the next section.

Contrary to the usual QNM models assumed in the literature, the amplitude $A(t,x)$ in (\ref{A_Integral}) is not necessarily given by a constant since the integration boundaries depend on $t$ and $x$, due to causality conditions. On the other hand, if the initial conditions are localized in a region smaller than $t-|x|$, then that region will determine the integration boundaries and $A$ will reach a constant for sufficiently large $t-|x|$. For example, if we had initial conditions with compact support like a  Gaussian: $\psi_0(x)=A_i\exp\{ - (\alpha x)^2/4\}$ and $\dot{\psi}_0=0$, we would obtain:
\begin{align}
\label{AA}
    A(t,x)&= A_i \sqrt{\pi}\frac{V_0}{\alpha}e^{\left(\frac{V_0}{2\alpha}\right)^2}\left[\text{Erf}\left(\frac{V_0}{2\alpha}-\frac{\alpha}{2} (t-|x|)\right)\right.\nonumber\\
    & \left. - \text{Erf}\left(\frac{V_0}{2\alpha} \right)\right],
\end{align}
where $\text{Erf}(y)$ is the Error function, which approaches $\pm 1$ when $|y|\gg 1$. This means that, due to causality, there will be a transitional period of $(t-|x|)$ in which the amplitude $|A(t,x)|$ will be growing towards a constant, as more of the signal has enough time to get in causal contact with the potential and reach the observer. We then emphasize that an evolving amplitude is not a sign of linear perturbation breaking, but it instead provides information about the shape of the initial conditions around the potential peak. This amplitude evolution was discussed in \cite{Andersson:1996cm}, and also illustrated in a toy example in \cite{Szpak:2004sf}. From this example, we also see that depending on the value of $V_0/\alpha$, the QNM amplitude reached asymptotically will not necessarily be of the same order of magnitude as the initial field value amplitude at the peak, $A_i$, unless $V_0/\alpha\sim \mathcal{O}(1)$. For instance, as $\alpha \rightarrow \infty$, the Gaussian initial condition will become narrower and there will be less signal available to reach the potential and observer, and one will obtain $A(t,x)\rightarrow 0$. Whereas for $\alpha\rightarrow 0$, there will be more signal available but there will be a limit anyway for any given point $(t,r_*)$ due to causality. Indeed, let us consider now  extended initial conditions (analogous to the $\alpha\rightarrow 0$ limit in the Gaussian initial condition example) so that $\psi_0$ and $\dot{\psi}_0$ are effectively constants in a region of size $2(t-|x|)$, then the amplitude would be:
\begin{equation}
\label{AAc}
    A(t,x) \approx \frac{4}{V_0} \left[e^{\frac{V_0}{2}(t-|x|)}-1\right]\left[\dot{\psi}_0-\psi_0\frac{V_0}{2}\right]. 
\end{equation}
After replacing this result into Eq.\ (\ref{Psi1_GralAmplitude}), we will obtain a QNM-like solution with a constant amplitude, in addition to a $(t,x)-$independent term that appears because the exponential in Eq.\ (\ref{Psi1_GralAmplitude}) cancels out the exponential term in Eq.\ (\ref{AAc}). This constant term illustrates the fact that additional non-QNM solutions may come from $G_Q$ in order to satisfy the initial conditions. 

Finally, we emphasize that, due to the integration limits in Eq.\ (\ref{A_Integral}), in an analogy with a Schwarzschild black hole, we expect the linear QNMs to be generated around the potential peak. 
On the contrary, the free waves in $\Psi_F$ are generated away from the potential peak, and their profiles are expected to mostly depend on the initial conditions near the horizon and at infinity.

Next, let us discuss the second-order solution. In analogy to the sources in Eq.\ (\ref{Source_Infinity}), let us assume a simple model where the quadratic source is given by:
\begin{equation}
    S^{(2)}(t,x)=C_s \frac{\Psi^{(1)2}_Q(t,x)}{(1+|V_0 x|)^2},\label{Quad_source}
\end{equation}
where $C_s$ is some arbitrary source constant and $\Psi^{(1)}_Q$ is in Eq.\ (\ref{Psi1_GralAmplitude}) (we consider only the first line corresponding to the QNM solution) with an exact constant amplitude $A$. As mentioned in Subsec.\ \ref{sec:QNM_Eqn}, it is important to add the $1/x^2$ suppression to the source, to avoid divergences in $\Psi^{(2)}$ in the limit of $V_0 x \rightarrow \infty$. The second-order solution with vanishing initial conditions can then be calculated as:
\begin{equation}\label{Psi2_Integral_DeltaV}
    \Psi^{(2)}= C_s A^2\int d\bar{t} d\bar{x} \; G(t,x|\bar{t},\bar{x}) \frac{e^{-V_0(\bar{t}-|\bar{x}|)}}{(1 + |V_0 \bar{x}|)^2}\Theta(\bar{t}-|\bar{x}|)\Theta(\bar{t}),
\end{equation}
which will have two contributions $\Psi^{(2)}_F$ and $\Psi_Q^{(2)}$ coming from $G_F$ and $G_Q$, respectively. We emphasize that $\Psi^{(2)}_F$ has no relation to $\Psi^{(1)}_F$ in these calculations, since $\Psi^{(2)}_F$ will be actually generated from $\Psi^{(1)}_Q$ due to the source (\ref{Quad_source}). The only commonality between $\Psi^{(1)}_F$ and $\Psi^{(2)}_F$ is that they are both propagated with the green's function $G_F$.
 
\begin{figure}[h!]
\centering
\includegraphics[width = 0.40\textwidth]{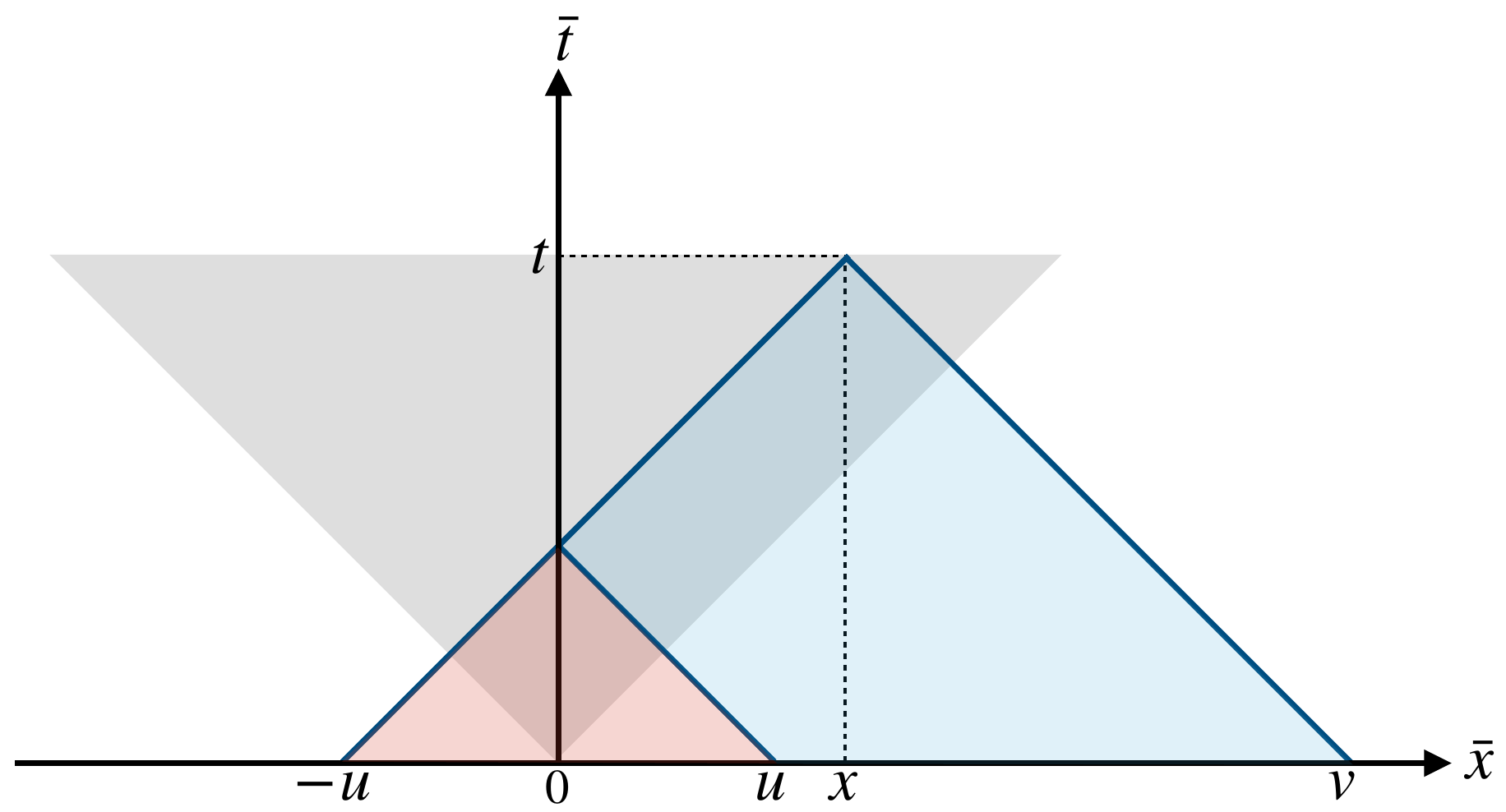}
 \caption{Support of the QQNM integrand for an observer at $x>0$ at a time $t$. The blue, red and grey shaded regions indicate the support of $G_F$, $G_Q$ and the quadratic source $S^{(2)}\sim \Psi^{(1)2}_Q$, respectively. 
 In the regions where the shades overlap is where $\Psi_F^{(2)}$  and $\Psi_Q^{(2)}$ have support.  Here, $u=t-x$ and $v=t+x$.}
 \label{Gsupport}
\end{figure}

Due to the step functions coming from the quadratic source and Green's function, the integrand in (\ref{Psi2_Integral_DeltaV}) will have support in a finite region of the $(\bar{t},\bar{x})$ space, which is illustrated in Fig.\ \ref{Gsupport}.
We emphasize that for an observer at $x>0$, $G_F$ always has support only for $\bar{x}>0$, regardless of the source. This makes sense given that $G_F$ describes free waves traveling ``directly'' to the observer without interacting with the potential. Therefore, we conclude that at first and second order, $G_F$ does not allow signals from one side of the potential barrier to be transmitted to the other side. However, from Fig.\ \ref{Gsupport}
we see that $G_Q$ has support in a region of positive and negative $\bar{x}$---yet always limited by $u$, regardless of the source. In the case of a source depending on $\Psi^{(1)}$, the region contributing to $\Psi^{(2)}_Q$ will be additionally limited by the support of the source.

It is easiest to describe the support of Fig.\ \ref{Gsupport} in terms of  $\bar{u}=\bar{t}-\bar{x}$ and $\bar{v}=\bar{t}+\bar{x}$ variables. For  $x>0$,  the integration limit of  $\Psi^{(2)}_F $ would be between $0<\bar{u}<u$ and $u<\bar{v}<v$, and for $\Psi^{(2)}_Q $ between  $0<\bar{u}<u$ and $0<\bar{v}<u$.
Using these limits of integration we obtain the following second-order solution:
\begin{align}
    &\Psi_F^{(2)}  = \frac{C_s A^2}{V_0^2}e^{-2} e^{-V_0 u} \left[ \text{Ei}\left(2\right) -e^{-2V_0x}\text{Ei}\left(2+2V_0x\right)  \right] \nonumber\\
    & + \frac{C_s A^2}{V_0^2}e^{-2} \left[ e^{-V_0 v}\text{Ei}\left(2+V_0v\right) - e^{-V_0 u} \text{Ei}\left(2+V_0u\right)\right], \label{PsiF_Delta}\\
     &\Psi_Q^{(2)} = 2\frac{C_sA^2}{V_0^2} e^{-V_0 u} \left[1-2 e^{-2} \text{Ei}(2) \right]\nonumber\\
     &   +  2\frac{C_sA^2}{V_0^2}e^{-\frac{V_0}{2} u}\left[ -1 + e^{-1}\text{Ei}(1)\right]\nonumber\\
     & + 2\frac{C_sA^2}{V_0^2}\left[2 e^{-2}e^{-V_0 u} \text{Ei}(2 + V_0 u ) - e^{-1}e^{-\frac{V_0}{2} u}\text{Ei}\left(1+  V_0 \frac{u}{2} \right) \right],\label{PsiV_Delta}
\end{align}
which also include the same causality condition $\Theta(t-|x|)$ as the linear QNM solution $\Psi^{(1)}_Q$, but it has been omitted in these expressions for compactness. In Eqs.\ (\ref{PsiF_Delta})-(\ref{PsiV_Delta}) we have introduced the exponential integral function---Ei---defined as:
\begin{equation}
     \text{Ei}(y)= -\int_{-y}^{\infty}d\bar{y}\; e^{-\bar{y}}/\bar{y}.
\end{equation}
For $x<0$, the solutions $\Psi_F^{(2)} $ and $\Psi_Q^{(2)} $ have the same expressions as in Eqs.\ (\ref{PsiF_Delta})-(\ref{PsiV_Delta}) making the replacement $x\rightarrow -x$, and hence $u\leftrightarrow v$. In obtaining these results, it was crucial to include all the causality conditions of the quadratic source and Green's function, otherwise the integrals for $\Psi^{(2)}$ would have diverged. 

More generally, from these results we first see that $\Psi^{(2)}_F$ has two contributions. The first line of Eq.\ (\ref{PsiF_Delta}) has a temporal evolution that goes as twice the linear QNM frequency. This line has the naive expected behaviour discussed in Subsec.\ \ref{sec:angle_time}, but notice that it has a non-trivial spatial evolution on its amplitude due to the terms depending on $v-u=2x$. Nevertheless, asymptotically for $V_0 x\rightarrow \infty$, we find that $e^{-V_0 (v-u)}\text{Ei}(2+V_0(v-u))\approx e^2/(2V_0 x)$ which vanishes and thus the first line of Eq.\ (\ref{PsiF_Delta}) describes a QNM term with an asymptotically constant amplitude. The second line of Eq.\ (\ref{PsiF_Delta}) does not have a typical oscillatory behaviour. For instance, for $V_0u, V_0v \gg 1$ we find that:
\begin{equation}
    e^{-V_0 v}\text{Ei}\left(2+V_0v\right) - e^{-V_0 u} \text{Ei}\left(2+V_0u\right) \approx \frac{e^2}{2+V_0 v}- \frac{e^2}{2+V_0 u}, \label{PsiF_Tail}
\end{equation}
which decays polynomially with time and distance. This tail was initially discussed in \cite{Okuzumi:2008ej}, where it was found that it appears due to the long-range behavior $1/x^2$ of the quadratic source and it is generated in asymptotically flat regions instead of near the potential. Indeed, if the source did not have a $1/x^2$ decay, we would have not obtained polynomial solutions in time.

Next, let us discuss the solution $\Psi^{(2)}_Q$ obtained in Eq.\ (\ref{PsiV_Delta}). In the first line we again see a term that behaves as twice the linear QNM frequency. Notably, the second line contains a term that behaves exactly like the linear QNM, which appears due to the presence of this linear frequency in the Green's function. 
In practice though, this second line is indistinguishable from the linear QNM which has arbitrary initial conditions. 
Next, the third line in  Eq.\ (\ref{PsiV_Delta}) describes again a power-law tail. For $V_0 u\gg 1$ we get:
\begin{align}
    &2 e^{-2}e^{-V_0 u} \text{Ei}(2 + V_0 u ) - e^{-1}e^{-\frac{V_0}{2} u}\text{Ei}(1+  V_0 u/2 )\approx \nonumber\\
    & -\frac{2}{(V_0 u)^2}\left[1 + \frac{2}{V_0 u} \right].
\end{align}
Comparing to Eq.\ (\ref{PsiF_Tail}), this tail coming from $\Psi^{(2)}_Q$ is subdominant at future null infinity. 
As highlighted in \cite{Okuzumi:2008ej}, these tails make perturbation theory break down at some point when $\Psi^{(2)}\simeq \Psi^{(1)}$ and, in that case, higher-order nonlinearities must also be included as well as possible first-order tails for extended potentials. 

We notice that $\Psi_F^{(2)}$ and $\Psi_Q^{(2)}$ end up having  comparable amplitudes in this toy model, even though the integration regions in Fig.\ \ref{Gsupport} are very different for $G_F$ and $G_Q$. This happens because the source decays with $|x|$ and $u$, which means the source emitted near the potential peak and around the moment of the merger is what mostly contributes to the solution, regardless of whether the signal interacted with the potential or not. In contrast, if the source did not have a $1/x^2$ suppression, we would find that $\Psi^{(2)}_F\rightarrow\infty$ and $V_0 x\rightarrow \infty$ due to the larger integration region contributing importantly. However, we would not obtain any divergent term in $\Psi^{(2)}_Q$ since the Green's function $G_Q$ and the source decay exponentially with $t-|x|$ and effectively limit the integration region of $\Psi^{(2)}_Q$ to $V_0(t-|x|)\lesssim 1$ anyway.

Furthermore, the fact that $\Psi^{(2)}_Q$ is a significant contribution to the total quadratic solution means that the QQNM signal detected by an asymptotically far observer still depends importantly on the source in a region of size $u$ around the potential---c.f.\ Fig.\ \ref{Gsupport}--- and not just to the right of the potential. In fact, in this model we find that $(1/2)\Psi_Q^{(2)}$ comes from the source at  $x<0$ whereas the other half comes from $x>0$. This is because the potential is symmetric around $x=0$, and hence it has equal transmission and reflection coefficients, and in addition the spatial profile of the source is also symmetric around $x=0$. 

Finally, if we assume that the QNMs dominate at intermediate times, compared to the polynomial tails and free waves, we can model the ringdown signal at these intermediate times as:
\begin{align}
    \Psi= \Psi_{1Q}+\Psi_{2Q}, \label{Toy_sol_1and2QNM}
\end{align}
where we have separated the terms that evolve with the linear QNM frequency, $\omega^{(1)}=-iV_0/2$ from those with the quadratic QNM frequency, $\omega^{(2)}=-iV_0$,
\begin{align}
   \Psi_{1Q} &=  e^{-\frac{V_0}{2}(t-|x|)}\left[ A +  2\frac{C_sA^2}{V_0^2}\left( -1 + e^{-1}\text{Ei}(1)\right)\right],\\
    \Psi_{2Q}&= \frac{C_s A^2}{V_0^2} e^{-V_0(t-|x|)-2} \left[2e^{2} -3\text{Ei}\left(2\right) -e^{-2V_0 |x|}\text{Ei}\left(2+2V_0|x|\right)  \right].
\end{align}
We emphasize that while $\Psi_{2Q}$ is a purely second-order perturbation, $\Psi_{1Q}$ contains both first and second-order perturbations now. In particular, if $A\ll V_0^2/C_s$ then $ \Psi_{1Q}$ will dominate the total $\Psi$ signal, but if $A\gg V_0^2/C_s$ then $ \Psi_{1Q}\sim \Psi_{2Q}$ for $t\sim |x|$, and $ \Psi_{1Q}\gg \Psi_{2Q}$ for $t\gg |x|+1/V_0$. In this example then, the linear QNM frequencies always determine a major/dominant contribution to the signal. 

Also notice that $\Psi_{1Q}$ and $\Psi_{2Q}$ satisfy the expected QNM boundary conditions, and locally propagate away from the potential in the limit of $V_0|x|\gg 1$, but their amplitudes are sensitive to the initial conditions and quadratic source on both sides of the potential. This local propagation behavior is the same one that we will find for a Schwarzschild black hole  in the eikonal limit in Sec.\ \ref{QNM_space}. This means that once the QNMs have been generated, the ones inside the light ring will propagate to the black hole and become unobservable.

Even though this toy example was extremely simple, the qualitative properties of its Green's functions are similar to those of a Schwarzschild black hole, and thus it allowed us to confirm basic features of the general solutions discussed in Subsec.\ \ref{secondGenerate}. However, certain differences are expected, including the obvious fact that there was not $G_B$ function in this toy model. For instance, the Zerilli and Regge-Wheeler potentials are not symmetric around their peaks, and their transmission and reflection coefficients may not be equal and will generically depend on the frequency of the quadratic QNM present in the source. This may introduce a preference for sources that come from $x<0$ or from $x>0$ to reach an asymptotic observer, and possibly play a role in determining how large nonlinearities are in observations. Relatedly, it is not clear whether $\Psi^{(2)}_F$ and $\Psi^{(2)}_Q$ will have  comparable amplitudes. 
In addition, since we make an angular decomposition into spherical harmonics, the linear amplitude in $\Psi_{1Q}$ may not be directly related to the quadratic amplitude in $\Psi_{2Q}$ for a given harmonic. As exemplified in Subsec.\ \ref{sec:angle_time}, this is the case of a $(\ell=4,|m|=4)$ harmonic, whose linear amplitude $A_{44}^{(1)}$ can be unrelated to the quadratic amplitude that mostly comes from 
the linear $(\ell=2,|m|=2)$ mode and hence scales as $~A^{(2)2}_{22}$.

Another difference is that the causality conditions of the Green's function and source that appeared as step functions in this toy model, will become smoother functions in a Schwarzschild background \cite{Szpak:2004sf}, and thus there may be a larger region of space around the light ring determining the amplitudes of the QNMs. All of these complications of a Schwarzschild black hole will have to be explored in more detail with the combination of numerical calculations in the future.

Finally, even within this toy model, there are extended analyses to deepen our intuition and understanding on the generation of QNMs. In particular, we assumed that the source was solely given by $\Psi^{(1)}_Q$ and ignored the effect of $\Psi^{(1)}_F$. However, generically the source should contain both parts. The importance of $\Psi^{(1)}_F$ will be fully dependent on the initial conditions but it will likely excite additional solutions with the linear QNM frequencies. Future investigations on realistic initial conditions will help discern the role of $\Psi^{(1)}_F$.
In addition, we could have also taken into account the fact that the amplitude of the linear QNM is not always constant, and analyzed induced variations in the amplitude of the quadratic solution. However, in the regime of a slow time variation in $A$, compared to $1/V_0$, we expect to have the same quadratic QNM result, now with an amplitude $A^2$ that includes a slow time drift at leading order.


\section{Local QQNM behavior}\label{QNM_space}
In this section we analyze the local radial behaviour of the QQNMs and confirm that not all of the waves travel to asymptotic observers, since in the eikonal limit the signals generated inside the light ring travel back to the black hole.

Let us consider Eqs.\ (\ref{Psi_Eq})-(\ref{Q_Eq}) for $j=2$. Due to the similarities between these two equations, all the qualitative results will be the same for both, and thus from now on we drop the odd and even superscript in $\Psi^{(j)}$.
In order to obtain analytical solutions that help gain intuition on the problem, we use the WKB approach, in analogy to what has been performed for linear QNMs in the past \cite{1985ApJ...291L..33S, 1987PhRvD..35.3621I}. 

In this section, we will not consider quadratic perturbations that have power-law behaviour or that behave as the linear QNMs. Instead, we only analyze the particular solutions with QQNM frequencies that are an addition or subtraction of two linear QNM frequencies.

\subsection{Asymptotic regime}\label{sec:WKB_asymptotic}
Due to the nearly constant shape of the potential towards the horizon and spatial infinity, one can use the WKB formalism to obtain asymptotic solutions to the equations of motion. In particular, a linear QNM with spherical harmonic number $\ell$ and eigenfrequency $\omega^{(1)}=\omega$ will have no source on its equation, and thus its asymptotic solution will be of the form \cite{1985ApJ...291L..33S}:
\begin{equation}\label{Linear_WKB}
    \Psi^{(1)}(r_*)\propto U(r_*;\omega,\ell)^{-1/4}e^{\pm i \int \sqrt{U(r_*;\omega,\ell)}dr_*},
\end{equation}
with a proportionality constant fixed by initial conditions. Here, we have defined the total radial potential $U(r_*;\omega,\ell)=\omega^2-V(r_*,\ell)\approx \omega^2$ as $r_*\rightarrow \pm \infty$, where $V$ can be $V_Z$ or $V_{RW}$. Note that this WKB solution holds when $U$ evolves slowly with radius (and hence represents a modulating amplitude) while the exponential term varies quickly. This is achieved when the phase $\sqrt{U}$ takes large values, which is the case in the eikonal limit, $\ell\gg 1$, since $\omega$ grows with $\ell$ according to linear theory calculations \cite{1985ApJ...291L..33S}. In this regime, the exponential term in Eq.\ (\ref{Linear_WKB}) varies quickly whereas the term $U^{-1/4}$ can be thought of as a slow varying amplitude.
In addition, the $\pm$ signs in the exponent of (\ref{Linear_WKB}) are chosen according to the QNM boundary conditions, that is, whether we are near the horizon and we have ingoing waves ($-$), or spatial infinity with outgoing waves ($+$). In particular, given the known asymptotic behaviour of the potentials $V_Z$ and $V_{RW}$, from Eq.\ (\ref{Linear_WKB}) we find that the linear QNMs behave as:
\begin{align}
&\Psi^{(1)} (r_*)\sim e^{+i\omega r_*}\quad  \text{for}\quad r_*\rightarrow +\infty,\label{LQNM_infinity}\\
& \Psi^{(1)} (r_*)\sim e^{-i\omega r_*} \quad \text{for}\quad r_*\rightarrow -\infty,\label{LQNM_horizon}
\end{align}
which are the usual boundary conditions that the quasi-normal waves satisfy. For concreteness, from now on, let us focus on the near horizon waves since the result obtained at spatial infinity will be analogous. 

Next, the linear solution (\ref{Linear_WKB}) will act as a source to the quadratic QNM variable $\Psi^{(2)}$. Without having an explicit expression of the quadratic source, in the WKB approximation we can still separate out the fast varying from the slow varying terms, given that we know the source to be a multiplication of background functions with two linear perturbations and their derivatives. In particular, the fast varying source terms can only come from the exponential piece in (\ref{Linear_WKB}). We then schematically express the quadratic equation of motion as:
\begin{align}
    &\Psi^{(2)''}(r_*)+U(r_*;\omega_2,\ell_2)\Psi^{(2)}(r_*)=S^{(2)}(r_*)\nonumber\\
    & =s^{(2)}(r_*) \exp\left\{- i \int dr_* \theta_{1\pm}(r_*;\omega,\ell;\omega',\ell')\right\},\label{QQNM_Eqn_WKB}
\end{align}
where $\theta_{1\pm}(r_*;\omega,\ell;\omega',\ell')$ can be:
\begin{align}
    \theta_{1+}\equiv & \sqrt{U(r_*;\omega,\ell)}+\sqrt{U(r_*;\omega',\ell')}, \nonumber\\
   \theta_{1-}\equiv & \sqrt{U(r_*;\omega,\ell)}-\left(\sqrt{U(r_*;\omega',\ell')}\right)^{*},
\end{align}
depending on whether the source does not include a conjugate $\Psi^{(1)}$ or it does (c.f.\ Eq.\ (\ref{QEqn_complex})). Here, due to the angular and temporal variable separation, we have assumed that only one pair of linear QNM solutions with $(\ell,m,\omega)$ and $(\ell',m', \omega')$ is sourcing a quadratic mode with given $(\ell_2,m_2,\omega_2)$, where $\omega_2$ can take two values: $\omega_2=\omega+\omega'$ when Eq.\ (\ref{QQNM_Eqn_WKB}) has a source with phase $\theta_{1+}$, or $\omega_2=\omega-\omega^{'*}$ when the source has $\theta_{1-}$. Similarly, we have the relationships $m_2=m\pm m'$ and $|\ell-\ell'|\leq \ell_2 \leq|\ell+\ell'|$. 

In addition, on the RHS of Eq.\ (\ref{QQNM_Eqn_WKB}) we assume $s^{(2)}$ to be a generally complex source function of $r_*$, that can also depend on the numbers $(\ell,m,\omega)$ and $(\ell', m',\omega')$ due to derivatives acting on the linear solutions $\Psi^{(1)}$ and due to the Clebsch-Gordan coefficients in Eq.\ (\ref{harmonic_product}). Nevertheless, this source function $s^{(2)}$  is expected to depend on finite maximum powers of $(\ell,\ell')$, as opposed to the exponential in (\ref{Linear_WKB}).
As a result, $s^{(2)}$ will evolve slowly in space compared to the exponential term in Eq.\ (\ref{QQNM_Eqn_WKB}) in the limit of $(\ell,\ell')\rightarrow\infty$, and is expected to approach zero in the asymptotic limit, according to Eqs.\ (\ref{Source_Infinity})-(\ref{Source_Horizon}).

Next, we use the WKB approach to solve  Eq.\ (\ref{QQNM_Eqn_WKB}). We introduce a small parameter $\eta$ that determines a scaling between slow and fast varying functions of radius. We thus rewrite Eq.\ (\ref{QQNM_Eqn_WKB}) as:
\begin{align}\label{Quadratic_WKB_EoM}
    &\eta^2 \Psi^{(2)''}(r_*) +U(r_*;\omega_2,\ell_2)\Psi^{(2)}(r_*)=s^{(2)}(r_*)\nonumber\\
    & \times \exp\left\{- \frac{i}{\eta} \int dr_*\theta_{1\pm}(r_*;\omega,\ell;\omega',\ell'))\right\},
\end{align}
where the total potential and source are expanded as:
\begin{align}
 U &= U_0(r_*) + \eta\; U_1(r_*) + \mathcal{O}(\eta^2),\\
    s^{(2)}&= s^{(2)}_0(r_*) + \eta\; s^{(2)}_1(r_*) + \mathcal{O}(\eta^2).
\end{align}
Given this hierarchy between the phase and the coefficients in the quadratic source, we  introduce the following WKB Ansatz for the quadratic QNM solution:
\begin{align}
    \Psi^{(2)}(r_*) = e^{-\frac{i}{\eta} \int dr_*\theta_2(r_*) } \left[ \Psi^{(2)}_0(r_*)+ \eta \Psi^{(2)}_1(r_*)+\mathcal{O}(\eta^2)\right],
\end{align}
which we replace into (\ref{Quadratic_WKB_EoM}) and obtain, at leading and sub-leading order in $\eta$, that
\begin{align}
 &\theta_2= \theta_{1\pm}, \label{theta_sol}\\
   &\Psi^{(2)}_0 =\frac{s^{(2)}_0} {[U_0-\theta_2^2]},\\
   & \Psi^{(2)}_1=\frac{s^{(2)}_1 + 2i\theta_{2} \Psi^{(2)'}_0-(U_1-i\theta_{2}')\Psi^{(2)}_0} {[U_0-\theta_2^2]}.
\end{align}
From these results we can express the leading-order WKB solution near the horizon as:
\begin{align}\label{WKB_asymptotic}
    \Psi^{(2)}(r_*)  \approx &\; S^{(2)}(r_*)/ \left\{ U(r_*;\omega_2,\ell_2)-\theta_{1\pm}^2\right\}.
\end{align}
This same expression will also hold at spatial infinity, with the difference that in that case the source goes as $S^{(2)}\propto \exp\{+i\int dr_*\,\theta_{1\pm}\}$ and thus $\theta_2=-\theta_{1\pm}$. Note also that the same functional form holds for the odd  and even quadratic perturbations.

Eq.\ (\ref{WKB_asymptotic}) allows us to obtain the asymptotic behaviour of $\Psi^{(2)}$, given the asymptotics of the source $S^{(2)}$ and of $(U(r_*;\omega_2,\ell_2)-\theta_{1\pm}^2)$. In particular, since at leading order  $U\approx \omega_2^2$ and $\theta_{1+}\approx \omega+\omega'$ and $\theta_{1-}\approx \omega-\omega^{'*}$, these leading-order expansions will cancel out and we will obtain that $(U(r_*;\omega_2,\ell_2)-\theta_{1\pm}^2)\propto r^{-2}$ at infinity, and $(U(r_*;\omega_2,\ell_2)-\theta_{1\pm}^2)\propto (r-2MG)$ near the horizon. For an asymptotically vanishing source $S^{(2)}$ with the same behaviour as the Zerilli and RW potentials (as assumed in Eqs.\ (\ref{Source_Infinity})-(\ref{Source_Horizon})) we would have at leading order that:
\begin{align}
&\Psi^{(2)}(r_*) \sim e^{+i\omega_2r_*}\quad  \text{for}\quad r_*\rightarrow +\infty,\label{QQNM_infinity}\\
& \Psi^{(2)}(r_*) \sim e^{-i\omega_2r_*} \quad \text{for}\quad r_*\rightarrow -\infty,\label{QQNM_horizon}
\end{align}
where we have used the fact that $\theta_{1\pm}\approx \omega_2$, which can be $\omega+ \omega'$ or $\omega- \omega^{'*}$. In either case, we see the same plane-wave behaviour as the linear QNM variable in Eqs.\ (\ref{LQNM_infinity})-(\ref{LQNM_horizon}), and thus the quadratic solutions (\ref{QQNM_infinity})-(\ref{QQNM_horizon}) satisfy the same boundary conditions of only ingoing waves at the horizon, and outgoing waves at spatial infinity as the linear QNMs. 
In addition, we see that if the source decayed slower asymptotically, e.g.\ $S^{(2)}\propto (r-2MG)^{0}$ near the horizon, then from Eq.\ (\ref{WKB_asymptotic}) we would find that $\Psi^{(2)}\propto e^{-i\omega_2 r_*}(r-2MG)^{-1}$ and would have a diverging power-law scaling. Similarly for any source that decays slower than $r^{-2}$ at spatial infinity. On the other hand, if the sources decayed faster than $(r-2GM)$ at the horizon or $r^{-2}$ at spatial infinity, the solution for $\Psi^{(2)}$ would also have a vanishing scaling. For this reason, the asymptotic choice in Eqs.\ (\ref{Source_Infinity})-(\ref{Source_Horizon}) is the more natural one, since in that case the variable $\Psi^{(2)}$ will be describing more directly the physical effects of nonlinearities such as the energy carried by quadratic QNMs, which does not diverge nor vanishes at the observer.

\subsection{Maximum of potential}

Next, we solve the quadratic QNM equation for $\Psi^{(2)}$ around the maximum of the potentials $V_Z(r_*,\ell)$ or $V_{RW}(r_*,\ell)$. In order to do that, we expand the potential around its maximum, which now does not necessarily vary slowly in radius compared to the spatial variations of $\Psi$. For both Schwarzschild potentials (\ref{VZ})-(\ref{VRW}), the maximum corresponds to the last circular stable photon orbit, at $\hat{r}\rightarrow 3MG$ (or $\hat{r}_*\approx 1.6 MG$)  when $\ell \rightarrow \infty$. The location of the maximum of the potential decays monotonically with $\ell$, and thus for small values of $\ell$ we will have that $\hat{r}>3MG$, but its variation with $\ell$ is slow and we will always be within $10\%$ of $3MG$ for any $\ell$.

Around the potential peak, we can make a second-order expansion in $(r_*-\hat{r}_*)$, such that 
both Zerilli and RW potentials take a simple parabolic form:
\begin{equation}\label{W_Taylor}
    U(r_*)\approx U(\hat{r}_*)+\frac{1}{2}U^{''}(\hat{r}_*)(r_*-\hat{r}_*)^2+\mathcal{O}\left((r_*-\hat{r}_*)^3\right),
\end{equation}
where $U^{''}(\hat{r}_*)=-V^{''}(\hat{r}_*)$ is the second-order derivative of the potential with respect to the tortoise coordinate, and can be calculated analytically from Eqs.\ (\ref{VZ})-(\ref{VRW}). We emphasize that in principle the expansion (\ref{W_Taylor}) is valid for $\Delta r \equiv (r_*-\hat{r}_*)/(MG)\ll 1$, regardless of the value of $\ell$. In particular, when $\ell \sim \mathcal{O}(1)$ then  $(MG)^4V^{''}\sim 10^{-2}$ and its higher derivatives are smaller, whereas for $\ell\rightarrow \infty$ we find that $(MG)^4V^{''}$ and all higher derivatives scale equally as $\ell^2$. Nevertheless, the neglected higher derivative terms, such as $(MG)^5V^{'''}$ get fractionally smaller with respect to $(MG)^4V^{''}$ as $\ell$ grows, and hence this approximation works better for large $\ell$ values. 

For this approximated potential, the linear QNM equation can be solved analytically. The solution satisfying the QNM boundary conditions has been found to given by \cite{1985ApJ...291L..33S}:
\begin{equation}
    \Psi^{(1)}(r_*)\propto H_n\left( z/\sqrt{2} \right) e^{-\frac{1}{4}z^2}; \quad z=(4k)^{\frac{1}{4}}e^{i3\pi/4}(r_*-\hat{r}_*),\label{Linear_ring_sol}
\end{equation}
with a complex proportionality constant depending on initial conditions. Here, we have introduced $k\equiv -V^{''}(\hat{r}_*;\ell)/2>0$ which is real and positive for a Schwarzschild black hole, and scales as $k\propto \ell^2$ in the eikonal limit. Also, the functions $H_n$ are the Hermite polynomials, which are polynomials of order $n$ containing only even (odd) powers of $z$  when $n$ is even (odd). The integer number $n\geq 0 $ describes the overtones of the linear QNM frequencies.
These solutions are valid for the ordinary QNMs with $\omega_R>0$, while an opposite sign for the exponent with $z^2$ in (\ref{Linear_ring_sol}) is obtained for mirror modes. From now on, without loss of generality, we focus on ordinary modes only.

In order to understand the solution (\ref{Linear_ring_sol}) better we analyze its limiting behavior. We first notice that $z$ is dimensionless and scales as $z\propto \sqrt{\ell}\Delta r$, so we can define an additional small parameter $\xi\equiv 1/\ell $, such that $\xi^{1/2} \ll \Delta r$ describes  
the limit of $z\rightarrow \infty$ as $\xi \rightarrow 0$.
While the solution (\ref{Linear_ring_sol}) is valid in general around the peak of the potential, this limit is of particular interest because it will give the solution that joins the previous WKB asymptotic expansion, and will allow us to use the WKB approach to solve for the quadratic solution later on. 
In this limit $z\rightarrow \infty$ as $\xi \rightarrow 0$, $\Psi^{(1)}$ becomes:
\begin{equation}
    \Psi^{(1)}(r_*)\propto (r_*-\hat{r}_*)^n e^{\frac{1}{2}i\sqrt{k}(r_*-\hat{r}_*)^2},\label{Linear_ring_away}
\end{equation}
which describes a wave with momentum $\sim \ell |r_*-\hat{r}_*|$ with $\ell \gg 1$, and a slow-varying amplitude modulation given by $(r_*-\hat{r}_*)^n$. 
Leaving aside this slow amplitude modulation and recovering the time dependence $\exp\{-i\omega t\}$ of the QNM solution, we then find that in the $|z|\gg 1$ limit  $\Psi^{(1)}$ has a time and radial fast evolution of the form:
\begin{equation}
    \Psi^{(1)}(r_*,t)\propto e^{-i \left[ \omega_R t - \frac{1}{2}\sqrt{k}(r_*-\hat{r}_*)^2\right]} e^{\omega_It},\label{Linear_ring_away_final}
\end{equation}
which describes waves propagating away from $\hat{r}_*$ when $\omega_R>0$, which is the case for the spectrum of the ordinary linear QNMs.
In particular, (\ref{Linear_ring_away_final}) describes waves propagating towards the horizon for $r_*< \hat{r}_*$, and towards spatial infinity for $r_*> \hat{r}_*$\footnote{Note that in this paper we have a different sign convention for outgoing and ingoing waves, when compared to \cite{1985ApJ...291L..33S}. This is why here the solution $\Psi^{(1)}\sim e^{+i\sqrt{k_1}(r_*-\hat{r}_*)^2/2} $ is the one that matches our QNM boundary conditions, whereas in \cite{1985ApJ...291L..33S} the authors choose $\Psi^{(1)}\sim e^{-i\sqrt{k_1}(r_*-\hat{r}_*)^2/2} $.}. We thus find that the region around the light ring determines the turning point for the linear QNM propagation direction in the eikonal regime. Putting this together with the previous WKB asymptotic solution, we conclude that high-frequency linear GWs generated inside the light ring propagate back to the black hole and become undetectable for asymptotic observers, whereas those generated outside the light ring become detectable.

Next, using this linear solution as a source, we calculate the quadratic solution around $\hat{r}_*$. In particular, we consider again the $|z|\gg 1$ regime and use the WKB approach. Let us make separate perturbative expansions in $\Delta r \ll 1$ and $\xi\ll 1$, using the hierarchy $\Delta r \gg \xi^{1/2}$. Note that while technically $\hat{r}_*$ (and hence $\Delta r$) depends on $\ell$ and that can introduce ambiguities on how to make these two separate expansions, in the $\ell\gg 1$ limit the $\hat{r}_*$ running with $\ell$ becomes negligible and we can simply approximate $\hat{r}\approx 3MG$ to a constant. In addition, the expansion on $\xi$ alone can become ambiguous in the quadratic QNM equation, since there are three different values of $\ell$ appearing: two coming from the linear QNMs in the source---$\ell$ and $\ell'$---and another one coming from the quadratic QNM itself---$\ell_2$. Here we assume that these three values are much larger than one and comparable, so that we can define a single perturbative parameter $\xi$ for the three harmonic values. 

We will first start by writing the approximate quadratic equation around the potential peak in the eikonal limit. In order to do that, we first Taylor expand the potential and the quadratic source in powers of $\xi$, and for each given power of $\xi$ we can then make a radial Taylor expansion around the maximum $\Delta r$. 
We start from Eq.\ (\ref{Psi_Eq}) and on its LHS we expand the function $U$, making use of the fact that  $\omega_2$ can be $\omega+\omega'$ or $\omega-\omega^{*'}$ for a pair of linear QNM frequencies $\omega$ and $\omega'$, whose analytical expressions are known in the eikonal limit \cite{1985ApJ...291L..33S, 1987PhRvD..35.3621I, Konoplya:2003ii}, and hence we know that $\omega^2$ and $V$ scale as $\sim \ell^2=\xi^{-2}$  at leading order:
\begin{equation}
    U(r_*;\omega_2,\ell_2)\approx \xi^{-2} \left[ u_{\xi 0}(r_*)  + \xi  u_{\xi 1}(r_*)+\cdots\right],\label{U_expanded}
\end{equation}
where the subscript $\xi n$ denotes the $n$-th order expansion in $\xi$. Here, $u_{\xi 0}$ and $u_{\xi 1}$ are functions of radius that do not depend on $\ell$. If we expanded these functions in a series of $\Delta r$, where we would find that their leading order term is of order $\Delta r^0$ with a next-to-leading order term $\Delta r^2$ (i.e.\ the radial evolution has a parabolic quadratic form around the peak, as expected):
\begin{equation}
    u_{\xi j}(r_*) = u_{0j}+ u_{2j}\Delta r^2, \label{u_radial_exp}
\end{equation}
where $u_{ij}$ are constant coefficients indicating the $i$-th power in $\Delta r$.
For the source on the RHS of Eq.\ (\ref{Psi_Eq}), we  do not make use of its specific functional form, nevertheless we know the source is formed as a product of two linear perturbations with arbitrary eigenfrequencies $\omega$ and $\omega'$, as well as harmonic and overtone numbers $(\ell,m, n)$ and $(\ell', m', n')$. Both of these linear perturbations could also appear with radial derivatives. In addition, these second-order terms will have a background coefficient that is expected to have a polynomial dependence on the radius $r$ (which can also be expressed as a power law dependence on $r_*$ around $\hat{r}_*$ when Taylor expanding), as well as a possible power-law dependence on the harmonic numbers $\ell$ and $\ell'$, and the eigenfrequencies $\omega$ and $\omega'$, appearing from possible angular and temporal derivatives acting on the linear perturbations, as well as the Clebsch-Gordan coefficients. 
In any case, we can schematically expand and separate out the source in terms of some slow and fast varying pieces as follows:
\begin{align}
   S^{(2)}(r_*, \xi) \approx & \xi^{q}\left[S_{\xi 0}(r_*)+\xi S_{\xi 1}(r_*) + \xi^2 S_{\xi 2}(r_*)+\cdots\right]\times \nonumber \\ 
   &e^{i\xi^{-1}\kappa_{\pm}(r_*-\hat{r}_*)^2/2},\label{Source_ring_zgg1}
\end{align}
where we have picked only two linear QNMs contributing to the source and, analogously to Subsec.\ \ref{sec:WKB_asymptotic}, their fast-varying phases (c.f.\ Eq.\ (\ref{Linear_ring_away})) are responsible for the fast-varying piece of the source in this eikonal limit, which lead to the constant exponent factor $\kappa_{\pm}\equiv \sqrt{k}/\ell\pm \sqrt{k'}/\ell'$. The $\pm$ sign in $\kappa$ determines whether one of the linear QNMs in the source was conjugated or not. This sign dependence shows explicitly that short-wavelength linear modes (i.e.\ with $\ell, \ell'\gg 1$) could source both short-wavelength quadratic modes (with $\ell+\ell'\gg 1$) and long-wavelength modes (with $|\ell-\ell'|\lesssim \mathcal{O}(1)$). In the eikonal approximation employed in this paper, we will assume that the source also has short-wavelength and thus we will require $|\ell-\ell'|\gg 1$.

Due to the fact that $\sqrt{k}\propto \ell$ in the limit of $\ell \gg 1$, we have made the $\ell$ dependence explicit in the source phase by introducing a $\xi^{-1}$ scaling in the exponent of Eq.\ (\ref{Source_ring_zgg1}), since the factors $k/\ell^2= U^{''}(\hat{r}_*;\ell)/2$ and $k'/\ell^{'2}= U^{''}(\hat{r}_*;\ell')/2$ are independent of $\ell$ and $\ell'$ \footnote{For consistency, $\sqrt{k}$ should also be expanded in leading and sub-leading powers in $\ell$, but in Eq.\ (\ref{Source_ring_zgg1}) we only keep the leading order one and that is why we can think of $\sqrt{k}/\ell$ as independent of $\ell$. We do this because the sub-leading terms in $\sqrt{k}$ can be reabsorbed into the functions $S_{\xi i}$, which are kept arbitrary here anyway.}. 
In addition, in  Eq.\ (\ref{Source_ring_zgg1}), we have introduced an additional arbitrary power of $\ell$, given by $\xi^{q}$ with $q$ some fixed number (that depends on e.g.\ background functions or angular derivatives appearing in the source), and we have truncated the $\xi$ expansion up to quadratic order.
Also, we have introduced the slow-varying radial functions $S_{\xi j}$ that describe the coefficients of each power of $\xi$ and are assumed to be independent of $\ell$\footnote{Note that the Taylor expansion of the source functions $S_{\xi j}$ is expected to have incremental powers of $\xi$ instead of $\xi^{1/2}$, even though the main variable we are using is $z\propto \xi^{-1/2}$. This is because the potentials and the linear solution $\Psi^{(1)}$ (see Eq.\ (\ref{Linear_ring_sol})) can be expanded on incremental integer powers of $\ell$. }. These functions can now be independently Taylor expanded in powers of $\Delta r$ as:
\begin{equation}
    S_{\xi j}(r_*) \approx \Delta r^{p_j}\left[c_{0 j}+  c_{1 j}\Delta r +  c_{2 j}\Delta r^2\right],
\end{equation}
where $p_j$ determines the dominant and lowest power of $\Delta r$ appearing in each source coefficient, while $c_{i j}$ describes the constant coefficient of each power $\Delta r^i$ in  $S_{\xi j}$. Here we have again truncated up to second order.  

Now that the source has a concrete form in our regime of interest, we can proceed to obtaining the particular solution to the QQNM equation by using the WKB approximation, and proposing an Ansatz that has analogous properties as the source:
\begin{align}
    \Psi^{(2)}\approx & \xi^{q+2}\left[\Psi_{\xi 0}(r_*)+\xi \Psi_{\xi 1}(r_*) + \xi^2 \Psi_{\xi 2}(r_*)\right]\times \nonumber \\ 
   &e^{i\xi^{-1}\kappa_{\pm}(r_*-\hat{r}_*)^2/2}
\end{align}
where we also expand
\begin{equation}
    \Psi_{\xi j}(r_*) \approx  \Delta r^{m_j}\left[\psi_{0 j}+  \psi_{1 j}\Delta r +  \psi_{2 j}\Delta r^2\right],
\end{equation}
with some powers $m_j$ and coefficients $\psi_{i j}$ to be determined.  Replacing the $\Psi^{(2)}$ Ansatz into the quadratic equation of motion with potential (\ref{U_expanded}) and source (\ref{Source_ring_zgg1}), we obtain at each order in $\xi$:
\begin{align}
   \Psi_{\xi 0}(r_*)=& \frac{- S_{\xi 0}}{(\kappa_{\pm}^2\Delta r^{2} - u_{\xi 0})},\\
   \Psi_{\xi 1}(r_*)=& \frac{-S_{\xi 1} + (u_{\xi 1}+i\kappa_{\pm})\Psi_{\xi 0}+2 i\kappa_{\pm}\Delta r \Psi_{\xi 0}^{'}}{(\kappa_{\pm}^2\Delta r^{2} - u_{\xi 0})},\\
   \Psi_{\xi 2}(r_*)=& \frac{- S_{\xi 2} + (u_{\xi 1}+i\kappa_{\pm})\Psi_{\xi 1} +2 i\kappa_{\pm}\Delta r \Psi_{\xi 1}^{'} +  \Psi_{\xi 0}^{''}}{(\kappa_{\pm}^2\Delta r^{2} - u_{\xi 0})}.
\end{align}
From here we also obtain the relation between the powers $m_j$ and $p_j$: $m_0=p_0$, $m_1=\min(p_1,p_0)$, and $m_2=\min(p_2,p_1, p_0-2)$. From these results it is straightforward to obtain the expressions for  $\psi_{i j}$ in terms of  $c_{i j}$ and $u_{i j}$, but we omit the explicit expressions here.

The main conclusion is that in the eikonal limit, the fast-varying radial and temporal dependence of the quadratic perturbation will then go as:
\begin{equation}
    \Psi^{(2)}(t,r_*)\propto e^{i(\sqrt{k}\pm \sqrt{k'})(r_*-\hat{r}_*)^2/2}e^{-i(\omega_{R}\pm \omega_{R}')t} e^{(\omega_I+\omega_I')t}\label{Psi2_zgg1_wave},
\end{equation}
with a proportionality function that evolves slowly with $r_*$. From the results of linear QNM perturbations in the eikonal limit \cite{1985ApJ...291L..33S} we know that $\omega_{R} \approx  \sqrt{V(\hat{r}_*,\ell)}$ which grows monotonically with $\ell$ and does not depend on the overtone $n$. In addition, we also find the same properties for $\sqrt{k}$. Therefore, in the quadratic solution (\ref{Psi2_zgg1_wave}) we see that both $(\sqrt{k}\pm \sqrt{k'})\propto (\ell\pm \ell')$ and $(\omega_{R}\pm \omega_{R}')\propto (\ell\pm \ell')$ will have the same sign, regardless of the harmonic and overtone numbers of the linear perturbations in the source.  This solution, in the large $|z|$ limit, thus describes a wave that propagates away from the potential peak, just as we confirmed for the linear QNM solution before. 

We notice that this result holds regardless of whether the source contained odd and/or even perturbations. This is because the linear QNM frequencies are the same for odd and even, and because both potentials $V_Z$ and $V_{RW}$ have the same form in the eikonal limit and thus both will lead to the same $k$ values. We also note that due to the simplicity of this calculation in the eikonal limit, one could iterate the result and obtain that higher-order perturbations will also describe waves propagating away from the potential peak in the eikonal limit. 

Finally, let us summarize the results of this section. In Subsec.\ \ref{sec:WKB_asymptotic} we showed that only very far away from the potential peak waves have a definite propagation direction, as expected from the QNM boundary conditions. However, in this subsection we generalize that result close to the potential peak as well. We thus conclude that waves in the eikonal limit have a definite propagation direction, and those located inside the light ring propagate to the black hole, whereas those located outside the light ring propagate to the observer. We emphasize though that the potential peak is not the exact location where the propagation direction turns over. Instead, there is a spatial region around the peak where the turnover happens, and the higher the frequency of the wave, the smaller the size of the turnover region. In particular, we expect the turnover to start happening when $z\sim \sqrt{\ell}\Delta r\sim 1$, which is when the approximation employed in this section breaks down. As a consequence, for small values of $\ell$ (that dominate GW signals from typical binary black hole mergers) there may be a considerably wide region around the potential peak where waves propagate in any direction, and this case will be investigated further in the future.

\section{Importance of nonlinearities}\label{energy}
In general, an understanding of nonlinearties will allow us to improve future ringdown models and maximize the science return from future GW events. 

First of all, if the inclusion of quadratic QNMs help improve ringdown models to earlier times (i.e.\ closer to the merger), then one can hope to use this early high signal-to-noise ratio (SNR) data available to detect more QNMs and include quadratic effects to avoid biases. When interpreting the results, one must take into consideration that some of the frequencies detected may not coincide with those of the linear QNM spectrum, and that would not indicate a violation of the no-hair theorem,  since these detections could correspond to GR nonlinear frequencies. This will happen if quadratic QNMs have a large enough amplitude to become detectable, as could happen for the $(\ell=4, |m|=4)$ angular harmonic (see estimations below).

Furthermore, by detecting quadratic QNMs, one can test the nonlinear dynamical predictions of GR. For any given pair of (linear/parent, quadratic/sourced) QNMs, one can confirm if their observed complex amplitudes satisfy the relationships expected from GR, which can be predicted from the linear QNMs (see \cite{Forteza:2022tgq} on how other work has proposed the use of linear QNM amplitude relationships to test GR). 

From the results of the previous sections we conclude that, after the signal has been generated, not all of the QNMs propagate towards an asymptotic observer but, during generation, the amplitude of the QNMs that do arrive at the observer are still influenced by the initial conditions close to the black hole horizon. Therefore, whether nonlinearities and, more precisely, quadratic QNMs can be observed in the ringdown close to the merger time, largely depends on what was the initial perturbation amplitude and its radial profile. 
In order to properly answer this question, realistic numerical simulations are necessary as well as the mathematical tools to connect near-BH physics to asymptotic physics (along similar lines to what has been done in \cite{Bhagwat:2017tkm}), both beyond the scope of this current paper. However, following \cite{Ioka:2007ak}, a simple dimensional analysis for equal-mass non-spinning black holes could be performed to show that the quadratic QNMs may have observable amplitudes, even if they are subdominant and perturbation theory works throughout the ringdown signal.

Let us assume that we are in a regime where perturbation theory works and continue using $\epsilon$ as the small expansion parameter (c.f.\ Eq.\ (\ref{h_expansion})). When analyzing observables, we are interested in the asymptotic behaviour of the fields in the limit of $r\rightarrow \infty$, so we  introduce a second perturbation parameter $\delta = G M/r$ and make an expansion of the metric at leading order in $\delta$, assuming the hierarchy $\delta \ll \epsilon$. We emphasize that the two expansions in $\epsilon$ and $\delta$ are independent since $\epsilon$ can be thought of as a perturbative expansion of the signal near the BH, whereas $\delta$ quantifies how far the signal is from the BH. Both perturbations  $\epsilon\Psi^{(1)}$ and $\epsilon^2\Psi^{(2)}$ will contribute to the metric at the same order in $\delta$ far from BH.

We then start by making a $\delta$ Taylor expansion of the metric perturbation $h_{\mu\nu}=\epsilon h^{(1)}_{\mu\nu}+\epsilon^2h_{\mu\nu}^{(2)}$ and reviewing its asymptotic behavior.
Since the metric is not gauge invariant and thus its asymptotic expression may take different forms, it is convenient to make the customary choice of asymptotically-flat gauge, in which the spatial components of $h_{\mu\nu}$ are transverse to the radial direction (for a radially-propagating wave) and will be the dominant terms in a $\delta$ expansion. Indeed, in this gauge, the leading-order asymptotic behaviour of the metric components in spherical coordinates is given by \cite{1979ApJ...230..870C}:
\begin{equation}
    h_{\theta \theta},h_{\theta\phi},h_{\phi\phi}\propto r; \;\; h_{t \theta}, h_{t \phi}, h_{r\theta},h_{r\phi}\propto r^{-1}; \;\; h_{tt},h_{t r},h_{rr}\propto r^{-2},\label{delta_exp}
\end{equation}
Here, we explicitly see that the dominant components are those coming from the angular indices, which are transverse but also asymptotically traceless since $h_{\theta\theta}+h_{\phi\phi}\propto r^0$. 
As a result, we only care about these angular components, which are determined by two physical degrees of freedom (DoFs), corresponding to the two polarizations carried by gravity in GR, which are in turn determined by the even and odd-parity fields ${}^{o,e}\Psi=\epsilon\; {}^{o,e}\Psi^{(1)} + \epsilon^2\; {}^{o,e}\Psi^{(2)}$. Note that in Cartesian coordinates, the asymptotic transverse traceless metric perturbations decay as $1/r$ and their amplitudes are estimated as \cite{Nagar:2005ea}
\begin{equation}
    h\sim \Psi/r,
\end{equation}
which, importantly, includes $\Psi^{(1)}$ and $\Psi^{(2)}$.
Therefore, the fields ${}^{o,e}\Psi$ can be interpreted as a proxy for the amplitude of the metric near the location of the source, in the radiation zone, and the expansion in $\epsilon$ separating $\epsilon\Psi^{(1)}$ from $\epsilon^2\Psi^{(2)}$ indicates the presence of a  hierarchy of amplitudes near the BH.
Next, we will see that these amplitudes are the ones that determine the energy carried by GWs and the relation between linear and quadratic QNMs.

At leading-order in $\delta$, and for appropriate definitions of the perturbations ${}^{o,e}\Psi$, the asymptotic power emitted in gravitational waves can be expressed as \cite{PhysRev.108.1063, PhysRevD.2.2141, Landau:1975pou}:
\begin{equation}\label{Power}
    G\frac{\partial E}{\partial t}= \sum_{\ell m n}\alpha_e(\ell)\bigg|\frac{\partial {}^{e}\Psi_{\ell m n}}{\partial t}\bigg|^2 +\alpha_o(\ell)\bigg|\frac{\partial {}^{o}\Psi_{\ell m n}}{\partial t}\bigg|^2,
\end{equation}
where we are using the $n$ index to label both the overtones of the linear QNMs as well as the the discrete modes in the quadratic QNM frequency spectrum for given $(\ell,m)$ harmonic numbers.
Here, we have additionally introduced the dimensionless functions $\alpha_{e,o}(\ell)$ that depend on the harmonic number $\ell$, but are assumed to be independent of $r$ and $t$. These functions have been introduced to generically describe  the arbitrary normalization of the functions $\Psi$ that has been varied in the past literature. As discussed in Sec.\ \ref{sec:QNM_Eqn}, the definition of the variables $\Psi^{(2)}$ involve some arbitrary choices, and here we choose them such that they encompass all the $\epsilon^2$ perturbations terms of the asymptotic transverse traceless metric, so that Eq.\ (\ref{Power}) holds. Indeed, previous works on quadratic perturbations obtain slightly different formulas due to their definitions of quadratic variables \cite{Gleiser:1995gx, Nakano:2007cj, Brizuela:2009qd}.

Separating the linear and quadratic contributions with different $\epsilon$ powers, we obtain
\begin{align}\label{Power_tot}
    &G\dot{E} \approx  \epsilon^2 G  \dot{E}^{(2)}\nonumber\\
    &+ 2 \epsilon^3\sum_{\ell m n} \alpha_e \Re\left[  \frac{\partial {}^{e}\Psi^{(1)}_{\ell m n}}{\partial t}\frac{\partial {}^{e}\Psi^{(2)*}_{\ell m n}}{\partial t} \right] + \alpha_o \Re\left[ \frac{\partial {}^{o}\Psi^{(1)}_{\ell m n}}{\partial t}\frac{\partial {}^{o}\Psi^{(2)*}_{\ell m n}}{\partial t}\right] \nonumber\\
    & +\epsilon^4  \sum_{\ell m n}\alpha_e(\ell)\big| {}^{e}\dot{\Psi}^{(2)}_{\ell m n}\big|^2 +\alpha_o(\ell)\big| {}^{o}\dot{\Psi}^{(2)}_{\ell m n}\big|^2 + \mathcal{O}(\epsilon^4),
\end{align}
where in the first line we have $E^{(2)}$ representing the energy coming from ${}^{o,e}\Psi^{(1)}$, the second line corresponds to the energy due to the mixing between first and second-order perturbations, and the third line includes purely second-order perturbations. As discussed in \cite{Brizuela:2009qd}, second-order perturbation theory only allows for a consistent energy calculation up to $\epsilon^3$ order, since third-order perturbations $\Psi^{(3)}$ will contribute to order $\epsilon^4$ to the energy and hence the third line in Eq.\ (\ref{Power_tot}) is technically incomplete. 

Recall that $\Psi^{(2)}$ includes both homogeneous solutions evolving with the linear QNM frequencies and particular solutions evolving with the quadratic QNM frequencies. From this feature, we expect the particular QNM solution to average out at order $\epsilon^3$ in the energy, after integrating the power in time. Thus, we expect the homogeneous solution of $\Psi^{(2)}$ and possible non-QNM solutions to mostly determine the energy at order $\epsilon^3$. 
We expect the particular solution of $\Psi^{(2)}$ to mostly contribute at order $\epsilon^4$ to the energy.

Next, we will use the fact that the energy depends quadratically on the QNM amplitude to estimate the importance of linear and quadratic perturbations in the strain.

Since the remnant black hole is what generates the ringdown GWs, the maximum energy GWs can radiate is $M$. Therefore, the energy radiated during the ringdown is usually quantified by the ringdown efficiency $\epsilon_{rd}$---the fraction of total black hole mass radiated in ringdown waves. The values can range between $\epsilon_{rd}\sim 0.8\%-3\%$ \cite{Flanagan:1997sx, Buonanno:2006ui, Berti:2007fi} depending on the binary mass ratio and spins. Given our assumption that perturbation theory works, most of the ringdown energy will be coming from the linear perturbation and thus we can estimate an order of magnitude of $E^{(2)}/M\sim 1\%$. 

As en example, in nearly equal-mass non-precessing quasi-circular BH binaries, the dominant QNM in the strain has $(\ell,m,n)=(2,2,0)$, and hence we can approximate the signal with a single QNM: ${}^{o,e}\Psi^{(1)}\approx  {}^{o,e}A^{(1)}_{220} \exp\{-i\omega^{(1)}_{220} t \}$. For this mode, the energy emitted due to linear perturbations is then estimated as:
\begin{equation}
    \frac{E^{(2)}}{M}\sim \left(\frac{\tilde{A}^{(1)}_{220}}{GM}\right)^2. \label{eq:220Amp}
\end{equation}
Here, $\tilde{A}^{(1)}_{220}$ is a proxy for the total linear amplitude at $t=0$ that includes both odd and even perturbations (and hence it is directly related to the amplitudes ${}^{o,e}A^{(1)}_{220}$). Note that here we have ignored the role of $\alpha_{o,e}$ since for $\ell \sim \mathcal{O}(1)$  we expect $\alpha_{o,e}\sim \mathcal{O}(1)$. 

From  Eq.\ (\ref{eq:220Amp}) we obtain  $\tilde{A}^{(1)}_{220}/(GM)\sim 10\%$. Next, let us estimate the dominant quadratic mode, which will be generated from the linear $(220)$ and will have harmonic numbers $(\ell=4,m=4)$ and frequency $\omega^{(2)}_{44}=2\omega^{(1)}_{220}$ associated.
Both $\Psi^{(1)}$ and $\Psi^{(2)}$ have the same units of $GM$, so a dimensional analysis would tell us that the amplitude of this quadratic QNM is of order
\begin{equation}
   \tilde{A}^{(2)}_{44} \sim (GM)^{-1}\tilde{A}^{(1)2}_{220}, \label{Lin_Quad_rel}
\end{equation}
which assumes that the coefficients in the quadratic source are of order $GM$, which is reasonable since both $\ell$ and the QNM frequency $\omega^{(1)}_{220}$ are present in the source and both are of order unity in this example.
In general, for large $\ell$ values additional non-unity factors may be expected.

From Eq.\ (\ref{Lin_Quad_rel}), we thus estimate that 
$\tilde{A}^{(2)}_{44}/(GM)\sim 1\% \sim 10\%\; \tilde{A}^{(1)}_{220}$.
Since for nearly equal-mass binary black holes, such as GW150914, the leading harmonic mode is $(2,2)$ and the next-to-leading harmonics, like $(4,4)$, have amplitudes that are about $1-10\%$ that of $(2,2)$, it is possible that the linear and quadratic QNM for the $(4,4)$ harmonic will have comparable or larger amplitudes near the moment of the merger. This estimate generally agrees with the numerical results found in \cite{London:2014cma, Mitman:2022qdl,Cheung:2022rbm} for the $(4,4)$ harmonic. As discussed in the introduction, next generation of GW detectors will have the ability to measure the $(4,4)$ harmonic even if its amplitude is a percentage of the dominant $(2,2)$ mode \cite{Berti:2016lat, Ota:2019bzl}, so we expect quadratic modes to be detectable in the future.

Note that in the estimation of Eq.\ (\ref{Lin_Quad_rel}),  Clebsch-Gordan coefficients should also appear, and hence affect the amplitude of $\Psi^{(2)}$. These coefficients take values close to $0.5$ or smaller, and suppress some linear harmonics to sourcing some quadratic harmonics. For instance, while the leading QQNM is expected to be in $(4,4)$, a sub-leading QQNM would be obtained from the product of linear modes such as $(2,2)\times (4,4)$. This combination of linear modes could source quadratic harmonics with $2\leq \ell \leq 6$  and $m= 6, 2$. However, the Clebsch-Gordan coefficient to the $(62)$ quadratic harmonic mode is always one or two orders of magnitude smaller than the rest. See Appendix \ref{app:CB} for a list of useful  Clebsch-Gordan coefficients and a further discussion on sub-leading quadratic modes. 

While in this discussion we assumed perturbation theory in $\epsilon$ up to second order, the energy expression (\ref{Power}) is generic asymptotically far, and can be made to include terms up to arbitrarily high powers of $\epsilon$. Since the total ringdown energy has to be smaller than $M$, we then expect $\tilde{A}^{(j)}/(GM)$ to always be smaller than unity even at higher orders in $\epsilon$. This may be a hint towards the more general validity of perturbation theory for the radiated GW signal, and will be investigated further in the future.

\section{Discussion}\label{conclusions}

The ringdown signal after the merger of two compact objects is typically analyzed using first-order perturbation theory, which is expected to work well some time after the merger, once any nonlinearities of the GWs have decayed.
Motivated by previous studies that have found linear perturbations to fit surprisingly well the GW signal even at the moment of the merger or slightly before, in this paper we analyze the role of nonlinearities by studying their qualitative physical properties regarding generation and propagation, and focusing particularly on second-order perturbations of a Schwarzschild black hole. 

Following earlier works, we use the Green's function approach to understand the generation of GWs. Since first and second-order perturbations (and higher order too) are calculated using the same Green's function, we find that they will all share certain common properties that we confirm by working out an explicit, fully analytical, toy example of quasi normal modes. First, we confirm that the ringdown signal (first and second order) can include quasi-normal modes (QNMs), as well as polynomial tails and arbitrary signals that depend on the initial conditions (sometimes referred to as the prompt response). 

Second, the causality constraints carried by the Green's function mean that at linear and nonlinear order the QNMs are generated in the region around the potential barrier peak, in the sense that QNMs are generated dynamically when signals are in causal contact with the potential peak (e.g.\ get reflected or transmitted by the potential barrier). This means that as time goes on, more signals interact with the potential and reach the observer, supporting a dynamical build-up picture of the QNM amplitudes. As a result, we find that the linear QNM amplitude can evolve in time before reaching a (typically assumed) constant amplitude, and thus we conclude that a time-evolving amplitude of QNMs is not necessarily a hint of perturbation theory breaking. Due to the same build-up picture, as time goes on, eventually the amplitude of the QNMs might be affected by the entire initial radial profile of the signal, both close and far from the black hole. However, it may also be plausible that, for practical purposes, the initial condition around the potential peak is the main factor determining the QNM amplitude. For this reason, a future numerical analysis on realistic spatial profiles for initial conditions may hold the key to answer the question of whether strong nonlinearities are present or not in observable ringdown signals. 

Third, we highlight previous results that show the linear QNM frequencies to be characteristic frequencies of the Green's function---as opposed to properties of just first-order perturbations. As a consequence, solutions with this linear frequency spectrum can be generated at any order in perturbation theory via the Green's function. In practice, this means that the amplitude of the QNMs with the linear frequency spectrum gets renormalized by receiving higher-order corrections, which may be a major reason why previous analyses found linear QNM frequencies to fit so well GW simulations even close to the merger. In addition, we confirm previous results on the presence of a new distinct higher-order spectrum of QNM frequencies, that are obtained by adding or subtracting linear QNM frequencies.

Furthermore, we analyze the local propagation behavior of linear and quadratic perturbations of a Schwarzschild black hole in the eikonal limit. We use the WKB approach to obtain the radial solutions to the linear and quadratic Zerilli and Regge-Wheeler equations. We find that both linear and quadratic perturbations effectively propagate away from the potential peak (approximately the light ring in this regime), which means that only the waves localized outside the light ring will propagate to an asymptotic observer and become detectable. Since, based on NR simulations, we expect most of the nonlinearities to be localized very close to the BH horizon, this result hints to why linear perturbation theory may work better than expected in describing asymptotically far signals. However, we emphasize that the local propagation behaviour obtained here only holds for high-frequency waves ($\ell\gg 1$), but typical GW signals are dominated by low-frequency waves, whose behavior will be analyzed further in the future.

We emphasize that further research is needed to fully understand the role and importance of nonlinearities in the ringdown. For instance, we assumed a perturbative approach that does not consider feedback effects to the black hole background (e.g.\ \cite{Sberna:2021eui}). However, due to the still large emission of GWs around the merger time, and due to GWs traveling back to the black hole, we expect the total mass and angular momentum of the black hole to initially evolve in time, and possibly affect the appropriate values of the QNM frequencies that one needs to use in ringdown models near the merger time. 
However, recent results have found linear QNMs to fit well the multipole moments of the dynamical horizon from a binary merger,  even when the horizon area still evolves significantly \cite{Mourier_2021}. It is still to be understood why this is the case, and whether it is just a numerical artifact due fitting time-limited signals with arbitrary number of linear QNMs.

In the future, we plan to generalize the qualitative analysis performed in this paper to Kerr black holes. However, we already expect various similarities. Linear and quadratic perturbations will satisfy the same radial Teukolsky equation with a vanishing source for the linear case and a  non-vanishing source for quadratic case \cite{Loutrel:2020wbw, Ripley:2020xby}, analogous to the Schwarzschild case. As a result, we again expect the same Green's function determining the linear and quadratic solutions, and hence propagating common properties to linear and nonlinear modes. Furthermore, the Green's function is expected to have similar qualitative properties to the one of a Schwarzschild black hole \cite{Yang:2013shb}. A WKB analysis of the solutions is also possible to perform \cite{PhysRevLett.52.1361,1993NCimB.108..991K, PhysRevD.41.374, Yang:2012he} since the Teukolsky equation can also be written in a Schr\"{o}dinger-like form with an effective potential barrier that peaks at a certain location \cite{1977RSPSA.352..381D}, just like for Schwarzschild black holes. 

Finally, we emphasize that the long-term goal of understanding nonlinearities holds great potential since it will allow us to improve ringdown models, improve the sensitivity of future no-hair theorem tests (by increasing the likelihood of detecting multiple QNM frequencies), and probe the dynamical non-linear behavior of gravity (by testing whether non-trivial physical effects get excited at higher-order that could modify the amplitude and frequency of quadratic QNMs expected in vacuum GR). For this reason, in this paper we argue that even if the nonlinearities were small during the entire ringdown signal, they will not necessarily be unobservable given the increased sensitivities of future GW detectors, and they will be worth exploiting. 

\section{ACKNOWLEDGMENTS}
We thank Leo Stein, Keefe Mitman and Sizheng Ma for useful conversations on the relevance of quadratic QNMs and their angular structure. M.L.\ also thanks the Benasque Science Center and the organizers of the 2022 workshop `New Frontiers of Strong Gravity' for seeding useful discussions. M.L.\ acknowledges NSF Grant No.\ PHY-1759835 for supporting travel to this workshop.
M.L.~was also supported by the Innovative Theory Cosmology fellowship at Columbia University.
L.H. was supported by the DOE DE-SC0011941 and a Simons Fellowship in Theoretical Physics.
\appendix

\section{Complex Variables}\label{app:complex}
Since the Einstein equations have at most two derivatives, the second-order equation of motion can be generally expressed as:
\begin{align}
    &G_{\mu\nu}^{(1)}(h^{(2)}) = L_{0\mu\nu}{}^{\alpha\beta}h^{(2)}_{\alpha\beta} + L_{1\mu\nu}{}^{\alpha\beta\gamma}h^{(2)}_{\alpha\beta;\gamma}+ L_{2\mu\nu}{}^{\alpha\beta\gamma\delta}h^{(2)}_{\alpha\beta;\gamma \delta},\\
    &S^{(2)}_{\mu\nu}(h^{(1)},h^{(1)}) = Q_{0\mu\nu}{}^{\alpha\beta\gamma\delta}h^{(1)}_{\alpha\beta}h^{(1)}_{\gamma\delta} +  Q_{1\mu\nu}{}^{\alpha\beta\gamma\delta\lambda}h^{(1)}_{\alpha\beta;\lambda}h^{(1)}_{\gamma\delta}\nonumber\\
    &+ Q_{2\mu\nu}{}^{\alpha\beta\gamma\delta\lambda\eta}h^{(1)}_{\alpha\beta;\lambda \eta}h^{(1)}_{\gamma\delta}+ Q_{3\mu\nu}{}^{\alpha\beta\gamma\delta\lambda\eta}h^{(1)}_{\alpha\beta;\lambda}h^{(1)}_{\gamma\delta; \eta},
\end{align}
where the tensors $L_i$ and $Q_i$ are formed using the background metric and its derivatives (and are therefore real), and the covariant derivatives are taken with respect to the background metric. From here we see that the real equations of motion can be expressed in terms of the complex variables by replacing:
\begin{align}
    &G_{\mu\nu}^{(1)}\left(\frac{1}{2}(h^{c(2)}+h^{c(2)*})\right)=\nonumber \\
    & S^{(2)}_{\mu\nu}\left(\frac{1}{2}(h^{c(1)}+h^{c(1)*}),\frac{1}{2}(h^{c(1)}+h^{c(1)*})\right)
\end{align}
which explicitly gives that:
\begin{align}
&L_{0\mu\nu}{}^{\alpha\beta}h^{c(2)}_{\alpha\beta} + L_{1\mu\nu}{}^{\alpha\beta\gamma}h^{c(2)}_{\alpha\beta;\gamma}+ L_{2\mu\nu}{}^{\alpha\beta\gamma\delta}h^{c(2)}_{\alpha\beta;\gamma \delta} + c.c. \nonumber\\
=& \frac{1}{2}\left[ Q_{0\mu\nu}{}^{\alpha\beta\gamma\delta}\left(h^{c(1)}_{\alpha\beta}h^{c(1)}_{\gamma\delta} +h^{c(1)*}_{\alpha\beta}h^{c(1)}_{\gamma\delta} \right)\right.\nonumber\\
& +  Q_{1\mu\nu}{}^{\alpha\beta\gamma\delta\lambda}\left(h^{c(1)}_{\alpha\beta;\lambda}h^{c(1)}_{\gamma\delta}+h^{*c(1)}_{\alpha\beta;\lambda}h^{c(1)}_{\gamma\delta} \right) \nonumber\\
    & + Q_{2\mu\nu}{}^{\alpha\beta\gamma\delta\lambda\eta}\left(h^{c(1)}_{\alpha\beta;\lambda \eta}h^{c(1)}_{\gamma\delta} + h^{*c(1)}_{\alpha\beta;\lambda \eta}h^{c(1)}_{\gamma\delta}\right)\nonumber\\
    &   \left. + Q_{3\mu\nu}{}^{\alpha\beta\gamma\delta\lambda\eta}\left( h^{c(1)}_{\alpha\beta;\lambda}h^{c(1)}_{\gamma\delta; \eta}+ h^{*c(1)}_{\alpha\beta;\lambda}h^{c(1)}_{\gamma\delta; \eta} \right)  \right] + c.c.
\end{align}
where $c.c.$ stands for complex conjugate. From here we explicitly see that the equation of motion for $h^{c(2)}$ can then be obtained as in Eq.\ (\ref{QEqn_complex}).

\section{Clebsch-Gordan Coefficients}\label{app:CB}
When calculating the Clebsch-Gordan Coefficients that relate the linear and quadratic QNMs, it is useful to keep track of the spin nature of the perturbation fields. In order to do that, it is best to work with the Newman-Penrose (NP) \cite{Newman:1961qr} and Geroch–Held–Penrose (GHP) \cite{1973JMP....14..874G} formalism, as it has been done for Kerr black holes in \cite{Bini:2002jx, Ripley:2020xby}. In the NP formalism one can express the ten metric perturbations in vacuum in terms of the five complex scalars $\Psi_0, \Psi_1, \Psi_2, \Psi_3, \Psi_4$, which have integer spins ranging from $-2$ to $+2$. $\Psi_4$ is the observably relevant field that determines the asymptotic behaviour of the metric, has spin $-2$, and is directly related to the Zerilli and Regge-Wheeler variables used throughout this paper (see explicit relationship in e.g.\ \cite{Nagar:2005ea}).
Therefore, we are interested in the quadratic source to $\Psi_4$, which will depend on all of the different metric components \cite{Campanelli:1998jv} and at most two derivatives acting in total. As a consequence, this source can contain the following spin-weighted spherical harmonic multiplications: ${}_{-2}Y_{\ell m}\times{}_{0}Y_{\ell' m'}$, ${}_{-1}Y_{\ell m}\times {}_{-1}Y_{\ell' m'}$, ${}_{-3}Y_{\ell m}\times{}_{1}Y_{\ell' m'}$ and ${}_{-4}Y_{\ell m}\times{}_{2}Y_{\ell' m'}$. 

If we consider a GW signal with leading linear mode from $(\ell,m)=(2,\pm 2)$ and a next-to-leading order mode $(4, \pm 4)$ (there could be others like $(3, \pm3)$ and $(3, \pm2)$ that we omit here for concreteness), then the relevant angular mixing coefficients that will appear in the quadratic source and determine the leading and next-to-leading quadratic QNMs are the following:

\begin{table}[h!]
\centering
\begin{tabular}{ |c|c|c| } 
 \hline
 $(-2,\ell,m)\times (0, \ell',m')$ & $(-2,\ell_2,m_2)$ & Angular Mixing \\ [0.5ex]  
 \hline\hline
 $(2,2)\times (2,2)$ & $(4,4)$ & $0.217641$  \\
 \hline
 $(2,2)\times (2,-2)$ & $(4,0)$ & $0.0260131$  \\
  & $(3,0)$ & $-0.119207$  \\
  & $(2,0)$ & $0.180224$  \\
 \hline
 $(2,2)\times (4,4)$ & $(6,6)$ & $0.197368$  \\
 \hline
 $(4,4)\times (2,2)$ & $(6,6)$ & $0.305761$  \\
 \hline
 $(2,-2)\times (4,4)$ & $(6,2)$ & $0.00887102$  \\
                      & $(5,2)$ & $0.053041$  \\
                      & $(4,2)$ & $0.12337$  \\
                      & $(3,2)$ & $0.132981$  \\
                      & $(2,2)$ & $0.0561946$  \\
 \hline
  $(4,4)\times (2,-2)$ & $(6,2)$ & $0.0137429$  \\
                      & $(5,2)$ & $-0.0410854$  \\
                      & $(4,2)$ & $-0.0424721$  \\
                      & $(3,2)$ & $0.206013$  \\
                      & $(2,2)$ & $0.217641$  \\
 \hline
\end{tabular}
\caption{Multiplication of linear modes with spins $s=-2$ and $s'=0$ that source a quadratic mode with spin $s=-2$.}\label{CB:n20}
\end{table}

\begin{table}[h!]
\centering
\begin{tabular}{ |c|c|c| } 
 \hline
 $(-1,\ell,m)\times (-1, \ell',m')$ & $(-2,\ell_2,m_2)$ & Angular Mixing \\ [0.5ex]  
 \hline\hline
 $(2,2)\times (2,2)$ & $(4,4)$ & $0.355406$  \\
 \hline
 $(2,2)\times (2,-2)$ & $(4,0)$ & $0.0424791$  \\
                      & $(2,0)$ & $-0.220728$  \\
 \hline
 $(2,2)\times (4,4)$ & $(6,6)$ & $0.353062$  \\
 \hline
 $(2,-2)\times (4,4)$ & $(6,2)$ & $0.015869$  \\
                      & $(5,2)$ & $0.0237206$  \\
                      & $(4,2)$ & $-0.0827594$  \\
                      & $(3,2)$ & $-0.208148$  \\
                      & $(2,2)$ & $-0.125655$  \\
 \hline
\end{tabular}
\caption{Multiplication of linear modes with spins $s=-1$ and $s'=-1$ that source a quadratic mode with spin $s=-2$.}\label{CB:n1n1}
\end{table}

\begin{table}[h!]
\centering
\begin{tabular}{ |c|c|c| } 
 \hline
 $(+1,\ell,m)\times (-3, \ell',m')$ & $(-2,\ell_2,m_2)$ & Angular Mixing \\ [0.5ex]  
 \hline\hline
 $(2,2)\times (4,4)$ & $(6,6)$ & $0.133445$  \\
 \hline
 $(2,-2)\times (4,4)$ & $(6,2)$ & $0.0059979$  \\
                      & $(5,2)$ & $-0.0448278$  \\
                      & $(4,2)$ & $0.121645$  \\
                      & $(3,2)$ & $-0.0786725$  \\
                      & $(2,2)$ & $-0.332452$  \\
 \hline
\end{tabular}
\caption{Multiplication of linear modes with spins $s=-3$ and $s'=+1$ that source a quadratic mode with spin $s=-2$.}\label{CB:p1n3}
\end{table}

\begin{table}[h!]
\centering
\begin{tabular}{ |c|c|c| } 
 \hline
 $(+2,\ell,m)\times (-4, \ell',m')$ & $(-2,\ell_2,m_2)$ & Angular Mixing \\ [0.5ex]  
 \hline\hline
 $(2,2)\times (4,4)$ & $(6,6)$ & $0.02359$  \\
 \hline
 $(2,-2)\times (4,4)$ & $(6,2)$ & $0.00106029$  \\
                      & $(5,2)$ & $-0.0126792$  \\
                      & $(4,2)$ & $0.0688127$  \\
                      & $(3,2)$ & $-0.222519$  \\
                      & $(2,2)$ & $0.470158$  \\
 \hline
\end{tabular}
\caption{Multiplication of linear modes with spins $s=-4$ and $s'=+2$ that source a quadratic mode with spin $s_2=-2$.}\label{CB:n4p2}
\end{table}

We emphasize that here we have used the fact that the GW signal for a mode $(\ell, m)$ is the same as that for $(\ell, -m)$ due to the presence of mirror modes. Here we see that the leading $(2,\pm 2)$ linear modes will not only source a $(4,\pm 4)$ quadratic harmonic but also some memory-like modes $(2,0)$, $(3,0)$, $(4,0)$ that do not oscillate but still decay exponentially in time. Note that the amplitude of the quadratic $(4,4)$ mode will be comparable to that of the $(2,0)$ and $(3,0)$ memory-like modes due to their comparable angular mixing coefficients. 

In addition, there can be various sub-dominant quadratic QNMs. As an example, the linear $(2,2)$ and $(4,4)$ QNMs can source a (6,2) mode, but its amplitude will tend to be suppressed compared to other modes due to the small angular mixing coefficient. Nevertheless, these same linear modes can also source a quadratic $(2,\pm 2)$ mode. If we consider a GW signal where the amplitudes near the merger time of the dominant linear (2,2,0) and (4,4,0) modes are related as $\tilde{A}^{(1)}_{440}\sim 10\%\, \tilde{A}^{(1)}_{220}$ and $\tilde{A}^{(1)}_{220}/(GM)\sim 10 \%$ (such as in nearly-equal masses quasi-circular binary mergers), then we could estimate the quadratic (2,2) amplitude $\tilde{A}^{(2)}_{22}$ as
\begin{equation}
    \tilde{A}^{(2)}_{22}\sim \mathcal{O}(0.1) (GM)^{-1}\tilde{A}^{(1)}_{440} \tilde{A}^{(1)}_{220}\sim 0.1\%\, \tilde{A}^{(1)}_{220},
\end{equation}
where the prefactor of $\mathcal{O}(0.1)$  comes from the various angular mixing coefficients that range from $0.06$ to $0.47$, depending on the spin of the linear modes present in the source.  Here we see that this quadratic mode in $(2,2)$ will have a sub-percent contribution to the total signal, so even though it typically decays slower than the first linear overtone $(2,2,1)$, it may not always have an observable impact. We emphasize though that these estimations are source-dependent and systems with high-mass ratios will have a different hierarchy of harmonic multipoles.

\bibliographystyle{apsrev4-1}
\bibliography{RefModifiedGravity.bib}

\end{document}